\documentclass[aps,preprint]{revtex4}

\usepackage{mathtools, bm}
\usepackage{amsmath,amssymb}  
\usepackage{textcomp} 
\usepackage[font=small]{caption}
\usepackage{mathrsfs}
\usepackage{xcolor}
 
\usepackage{graphicx}
\usepackage{subfigure}
\usepackage{amsmath}
\usepackage{xcolor}
\usepackage{color, soul}
\usepackage{tabularx}
\usepackage{wrapfig}
\newcommand{\beq}{\begin{equation}}
\newcommand{\eeq}{\end{equation}}

\newcommand{\rhos}{\rho_{s \perp}}
\newcommand{\bpe}{\beta_{\perp_e}}

\newcommand{\hx}{\hat{x}}
\newcommand{\hy}{\hat{y}}
\newcommand{\hz}{\hat{z}}

\newcommand{\bx}{\mathbf{x}}
\newcommand{\br}{\mathbf{r}}
\newcommand{\boldy}{\mathbf{y}}

\newcommand{\hbr}{\hat{\mathbf{r}}}

\newcommand{\Tpas}{T_{{0 }_{\parallel s}}}
\newcommand{\Tpes}{T_{{0 }_{\perp s}}}
\newcommand{\Tpae}{T_{{0}_{\parallel e}}}
\newcommand{\Tpee}{T_{{0}_{\perp e}}}

\newcommand{\hTpae}{\hat{T}_{{\parallel e}}}
\newcommand{\hTpee}{\hat{T}_{{\perp e}}}
\newcommand{\hTpas}{\hat{T}_{{\parallel s}}}
\newcommand{\hTpes}{\hat{T}_{{\perp s}}}

\newcommand{\tpas}{T_{\parallel s}}
\newcommand{\tpes}{T_{\perp s}}
\newcommand{\tpae}{T_{\parallel e}}
\newcommand{\tpee}{T_{\perp e}}

\newcommand{\apar}{A_{\parallel}}
\newcommand{\bpar}{B_{\parallel}}
\newcommand{\bpara}{B_{\parallel}}
\newcommand{\lapp}{\Delta_\perp}

\newcommand{\wapar}{ \widetilde{A}}
\newcommand{\wphi}{\widetilde{\phi}}
\newcommand{\hapar}{\hat{A}_{\parallel}}

\newcommand{\hphi}{\hat{\phi}}

\newcommand{\hbpar}{\hat{B}_{\parallel}}
\newcommand{\hxi}{\bar{\xi}_{in}}

\newcommand{\de}{\delta}
\newcommand{\nno}{\nonumber}

\newcommand{\byo}{B^{(0)}_{y}}
\newcommand{\coeff}{\frac{ g \Theta_e}{2 \alpha \lambda}}
\newcommand{\tpi}{\tau_{\perp_i}}

\newcommand{\hmua}{\hat{\mu}_{s}}
\newcommand{\hht}{\hat{t}}
\newcommand{\qa}{q_{s}}
\newcommand{\dwa}{d \mathcal{W}_s}

\newcommand{\hcalfa}{\hat{\mathcal{F}}_{{eq_s}}}

\newcommand{\hga}{\hat{g}_s}

\newcommand{\hvpar}{\hat{v}_\|}
\newcommand{\thea}{\Theta_s}
\newcommand{\bepes}{\beta_{\perp_s}}
\newcommand{\Guo}{\mathcal{G}_{{10_s}}}
\newcommand{\Gdo}{\mathcal{G}_{{20_s}}}
\newcommand{\Guu}{\mathcal{G}_{{11_s}}}
\newcommand{\Gdu}{\mathcal{G}_{{21_s}}}
\newcommand{\Guoe}{\mathcal{G}_{{10_e}}}
\newcommand{\Gdoe}{\mathcal{G}_{{20_e}}}
\newcommand{\Guue}{\mathcal{G}_{{11_e}}}
\newcommand{\Gdue}{\mathcal{G}_{{21_e}}}
\newcommand{\taups}{\tau_{\perp_s}}

\newcommand{\hpdf}{\hat{\mathsf{f}}_s}
\newcommand{\vtpa}{v_{{th }_{\parallel s}}}

\newcommand{\hvpa}{\hat{v}_{_{\parallel}}}
\newcommand{\vpe}{v_{_{\perp }}}
\newcommand{\hvpe}{\hat{v}_{_{\perp }}}
\newcommand{\hns}{\hat{N}_s}
\newcommand{\hus}{\hat{U}_s}

\newcommand{\hdfa}{\hat{f}_{s}}
\newcommand{\ns}{N_s}
\newcommand{\us}{U_s}

\newcommand{\hap}{\hat{A}_\parallel}
\newcommand{\bk}{\mathbf{k}}
\newcommand{\hbk}{\hat{\mathbf{k}}}

\newcommand{\hbrhos}{\hat{\boldsymbol{\rho}}_{\perp_s }}

\newcommand{\aal}{a_{s}}
\newcommand{\bps}{\beta_{\perp_s}}
\newcommand{\lbphi}{\bar{\mathcal{L}}_\phi}
\newcommand{\lbb}{\bar{\mathcal{L}}_B}
\newcommand{\lba}{\bar{\mathcal{L}}_A}

\newcommand{\kp}{k_{\perp}}

\begin{document}

\title{Impact of electron temperature anisotropy on the collisionless tearing mode instability in the presence of a strong guide field}

\author{C. Granier$^{1,2,*}$\thanks{corresponding author}, E. Tassi$^{1}$, D. Borgogno$^{2,3}$, D. Grasso$^{2,3}$}
\affiliation{$^1$ Universit\'e C\^ote d'Azur, CNRS, Observatoire de la C\^ote d'Azur, Laboratoire J. L. Lagrange, Boulevard de l'Observatoire, CS 34229, 06304 Nice Cedex 4, France\\
$^2$  Dipartimento di Energia, Politecnico di Torino, Torino 10129, Italy\\
$^3$ Istituto dei Sistemi Complessi - CNR and Dipartimento di Energia, Politecnico di Torino, Torino 10129, Italy\\
$*$ Corresponding author: camille.granier@oca.eu}

\date{\today }

\baselineskip 24pt

\begin{abstract}

We derive and analyze a dispersion relation for the growth rate of collisionless tearing modes, driven by electron inertia and accounting for equilibrium electron temperature anisotropy in a strong guide field regime.  For this purpose, a new gyrofluid model is derived and subsequently simplified to make the derivation of the dispersion relation treatable analytically. The main simplifying assumptions consist in assuming cold ions, neglecting electron finite Larmor radius effects, decoupling ion gyrocenter fluctuations and considering $\beta_{\perp_e} \ll 1$, with $\beta_{\perp_e}$ indicating the ratio between the perpendicular electron thermal pressure and the magnetic pressure exerted by the guide field. This simplified version of the gyrofluid model is shown to possess a noncanonical Hamiltonian structure. The dispersion relation is obtained by applying the theory of asymptotic matching and does not predict an enhancement of the growth rate as the ratio $\Theta_e$, between perpendicular and parallel equilibrium electron temperatures, increases. This indicates a significant difference with respect to the case of absent or moderate guide field. For an equilibrium magnetic shear length of
the order of the perpendicular sonic Larmor radius and at a fixed $\bpe$, we obtain that the tearing mode in the strong guide field regime gets actually weakly damped, as $\Theta_e$ increases. In the isotropic limit $\Theta_e=1$, the dispersion relation reduces to a previously known formula. The analytical predictions are tested against numerical simulations showing a very good quantitative agreement. We also provide a detailed discussion of the range of validity of the derived dispersion relation and of the compatibility among the different adopted assumptions.

\end{abstract}

\maketitle

\section{Introduction}  \label{sec:intro}

 Magnetic reconnection is a process believed to play a key role in many phenomena occurring in laboratory and astrophysical plasmas, such as magnetospheric substorms, coronal mass ejections and  sawtooth crashes in tokamaks \cite{Pri00,Bis00,Yam10}. 
In collisionless plasmas as, for instance, the magnetosphere and the solar wind, mechanisms alternative to collisional resistivity, are the main responsibles for the violation of the frozen-in condition allowing for reconnection of magnetic field lines. Electron inertia, becoming particularly relevant at the scale of the electron skin depth, can provide  one of such mechanisms. A further feature of collisionless plasmas is that they can exhibit particle distribution functions which are anisotropic with respect to the direction of the magnetic field. This can lead in particular to anisotropic temperature distributions. A natural question in the theory of collisionless plasmas concerns then the influence of temperature anisotropy on characteristic features of reconnection, such as the linear growth rate of the tearing mode. The influence of equilibrium temperature anisotropy on the reconnection growth rate has actually been the object of several studies \cite{For68,Che84,Shi87,Chi02,Kar04,Dau05,Mat08,Que10} carried out with kinetic and fluid approaches. Such studies agree with predicting that temperature (and in particular electron temperature) anisotropy, enhances the growth rate, meaning that the growth rate increases as the ratio between the perpendicular and parallel temperature increases, where perpendicular and parallel are referred to the direction of the equilibrium magnetic  field. Such studies, on the other hand, consider the case of absent or moderate magnetic guide field. The investigation of Ref. \cite{Shi87} also indicates that, when the amplitude of the guide field is increased from zero to $2.5$ times  the amplitude of the equilibrium field in the reconnection plane, the enhancement of the growth rate gets weakened.

 In this paper we carry out an analytical and numerical investigation of the influence of electron temperature anisotropy on the reconnection growth rate  in the opposite regime, i.e. the regime of strong guide field. 
 In our analysis we also assume the aforementioned electron inertia as responsible for the violation of the frozen-in condition. An effective tool, which we adopt here, for modelling inertial reconnection in the strong guide field regime, is provided by reduced gyrofluid models. Such models are derived from gyrokinetic equations, which namely assume a strong guide field ordering. This is also related to the choice of the perpendicular sonic Larmor
radius, $\rhos$, as characteristic length, which also implies that our study focuses on a microscopic current sheet, as opposed to the macroscopic case of magnetohydrodynamics (MHD) \cite{Del17}.
Such regime can be relevant  for a variety of plasmas in some regions of the solar wind, where the guide field is taken to be the mean magnetic field \cite{Sch09} and currents sheet are small. We remark that collisionless magnetic reconnection in the presence of a strong guide field and electron temperature anisotropy,  occurring in microscopic current sheets, was observed in the Earth's magnetosheath \cite{Pha18, Eas18}. In particular, Ref. \cite{Eas18} reports the observation of magnetic structures having a width of the order of the sonic Larmor radius, in the magnetosheath, in the presence of a strong guide field and anisotropic changes in ion and electron temperature.  
 
 In the context of magnetic reconnection, reduced gyrofluid models have been previously adopted, for instance in order to study the influence of ion Finite Larmor Radius (FLR) effects on the linear and nonlinear evolution of collisionless reconnection \cite{GCPP_2011,Com12}. However, such models, did not account for equilibrium temperature anisotropy, one of the essential features required for our present analysis. Therefore, we present here also the derivation of a new reduced gyrofluid model, which takes into account equilibrium temperature anisotropy, and which will be the basis for our derivation and analysis of the dispersion relation for the linear growth rate. The gyrofluid model we present here, in particular,  differs from similar gyrofluid models recently derived \cite{Tas19,Tas20}, as it assumes an isothermal closure referred to parallel and perpendicular particle temperatures.
 
  Although we derive a gyrofluid model accounting, in addition to equilibrium temperature anisotropy, also for FLR as well as finite $\beta$ effects (with $\beta$ indicating the ratio between the plasma equilibrium thermal pressure and the magnetic pressure associated with the guide field), the actual model adopted for the linear stability analysis will be a simplified Hamiltonian version of the general gyrofluid model. In particular, we will limit to the cold-ion limit, neglect the contribution of ion gyrocenter fluctuations as well as of electron FLR effects, and consider the regime $\bpe \ll 1$. These simplifying assumptions clearly reduce the range of applicability of our analysis but, on the other hand, make the problem amenable to a fully analytical treatment, the results of which we will also validate by means of numerical simulations. The analytical treatment will also make it possible to compare the resulting dispersion relation for the growth rate, with a previously derived analytical formula, obtained assuming similar simplifying assumptions but not accounting for temperature anisotropy \cite{Por91,Fit07,Fit10,Tas18}. An analysis based on a more refined version of the gyrofluid model requires a mostly numerical approach, and is part of our current work in progress.
 
  The paper is organized as follows. In Sec. \ref{sec:model} we present the model adopted for the linear analysis, describe the main underlying assumptions and review its Hamiltonian structure. Section \ref{sec:lintear} contains the derivation of the analytical dispersion relation for the growth rate, which is based on asymptotic matching theory.  The predictions based on the analytical dispersion relation are checked against numerical simulations in Sec. \ref{sec:num}. In Sec. \ref{sec:concl} we first summarize and discuss the conditions of applicability of the analytical dispersion relation and then we conclude, indicating also possible future developments. In  Appendix \ref{app:mod} we present the derivation of the new gyrofluid model, starting from a gyrokinetic system, and the reduction to its simplified form adopted for the linear analysis. In Appendix \ref{app:lim} we describe a technical step, required to show that the outer solution for the electrostatic potential, derived in Sec. \ref{ssec:outer}, satisfies the appropriate boundary condition.

\section{Model equations for the tearing stability analysis}  \label{sec:model}

We consider,  in a slab geometry equipped with Cartesian coordinates $x$, $y$ and $z$, a magnetic field $\textbf{B}$ whose approximate expression, in terms of dimensionless variables later defined in Eq. (\ref{norm}), is given by
\begin{equation} \label{magfield}
     \mathbf{B}(x,y,z,t) \approx  \mathbf{z}+ \bpar(x,y,z,t)\mathbf{z} + \sqrt{\frac{\bpe}{2}} \nabla \apar (x,y,z,t)\times \mathbf{z},
\end{equation}
where $t$ is the time coordinate, $\apar$ is the perturbation of the parallel component of the magnetic vector potential and $\bpar$ is the perturbation of the magnetic field in the parallel direction. In Eq. (\ref{magfield}), the first term on the right-hand side accounts for a  uniform guide field,  directed along the unit vector $\mathbf{z}$ and assumed to be strong, which implies $\bpar \ll 1$ and $\vert \nabla \apar \vert \ll 1$. The subscripts $\parallel$ and $\perp$ denote
components parallel and perpendicular to the direction of the guide field. The expression for $\mathbf{B}$ in Eq. (\ref{magfield}) is approximate as it is not divergence-free and represents the expression of the magnetic field at the first order in the fluctuations. The higher-order contributions, which guarantee $\nabla \cdot \mathbf{B}=0$, are neglected. Indeed, already in the gyrokinetic model \cite{kunz2015} which our model is derived from (Eqs. (\ref{1})-(\ref{4})), such terms only provide higher order contributions which are neglected.\\

We introduce the following  dimensionless variables:
\begin{equation} \label{norm}
\begin{split}
  &  x=\frac{\hat{x}}{\rhos}, \qquad    y=\frac{\hat{y}}{\rhos},  \qquad  z=\sqrt{\frac{\bpe}{2}}\frac{\hat{z}}{\rhos}, \qquad   t=\omega_{ci} \hat{t},  \\
  & N_e=\frac{\hat{N}_e}{n_0},  \qquad    U_e=\sqrt{\frac{\bpe}{2}} \frac{\hat{U}_e}{c_{s \perp}}, \qquad   \tpee = \frac{\hTpee}{\Tpee}, \qquad   \tpae = \frac{\hTpae}{\Tpae},\\
   & \phi=\frac{e \hphi}{T_{0_{\perp e}}}, \qquad   \bpar=\frac{\hbpar}{B_0},  \qquad \apar= \frac{1}{\rhos} \sqrt{\frac{2}{\bpe}} \frac{\hapar}{B_0},
\end{split}  
\end{equation}
where the dimensional  dependent and independent variables are denoted by the hat symbol.  We indicate with $\rhos$ the sonic Larmor radius based on the perpendicular electron temperature, which we also take as   characteristic length-scale of the perturbations. The homogeneous equilibrium state is characterized by a magnetic guide field of amplitude $B_0$, a constant and uniform density $n_0$  and can admit anisotropic temperatures  $\Tpes$, $\Tpas$, where the index $s$ denotes the particle species. We assume that the plasma consists of two species, electrons and single ionized ions respectively, so that the index $s$ can take only the two values $s = e$ and $s = i$. We denote by $\omega_{cs}= e B_0/ (m_s c)$ the cyclotron frequency of the species $s$, where $m_{s}$ indicates the mass of the particle,  $e$ is the proton charge and $c$ is the speed of light. The symbol  $c_{s \perp} =\sqrt{T_{0_{\perp e}} / m_i} $ denotes the sound speed based on the perpendicular temperature and $\bpe=8 \pi n_0 T_{0_{\perp_e}}/B_0^2$ indicates the ratio between the equilibrium perpendicular thermal electron pressure and the magnetic pressure. Note that, following a customary notation, the sonic Larmor radius $\rhos$ and the perpendicular sound speed $c_{s \perp}$, which are related by $\rhos=c_{s \perp}/\omega_{ci}$, are indicated with symbols containing the subscript $s$, although in these two cases such subscript does not refer to the particle species.  Finally, the independent variables ${N}_e$, ${U}_e$, $\tpee$ and $\tpae$ indicate electron gyrocenter fluctuations of density, parallel velocity, perpendicular and parallel temperature, respectively, whereas we denote with $\phi$ the fluctuations of the electrostatic potential. The latter independent variables are all functions of $x$, $y$, $z$ and $t$. Concerning the normalization of $\apar, U_e$ and $z$ in Eq. (\ref{norm}), we think it may be worth pointing out that, in comparison for instance with Ref. \cite{Tas18}, the factor $\sqrt{2/ \bpe}$ was introduced here in order to make the limit $\bpe \rightarrow 0$ more transparent in the derivation of the model equations from the gyrofluid model presented in Appendix \ref{app:mod}. This also explains the presence of the coefficient $\sqrt{\bpe /2}$ appearing in Eq. (\ref{magfield}).

The model equations adopted for the tearing stability analysis are
\beq \label{continuity}
\frac{\partial \lapp \phi}{\partial t} + [\phi, \lapp \phi] - [\apar, \lapp \apar] + \frac{\partial \lapp \apar}{\partial z}=0,
\eeq
\begin{equation}
\begin{split} \label{ohmslaw}
\frac{\partial}{\partial t} \left( \apar -  \frac{2\de^2}{\bpe} \lapp \apar \right) + \left[\phi, \apar  -  \frac{2\de^2}{\bpe} \lapp \apar \right]  - \frac{1}{\Theta_e} [\lapp \phi, \apar] + \frac{\partial }{\partial z}\left( \phi - \frac{\lapp \phi}{\Theta_e} \right)=0.
\end{split}
\end{equation}
where the operator $[ \, , \, ]$ is the canonical Poisson bracket defined by $[f,g]=\partial_x f \partial_y g - \partial_y f \partial_x g$, for two functions $f$ and $g$. The model (\ref{continuity}) - (\ref{ohmslaw}) is given in its three-dimensional (3D) form. However, for the analysis of the tearing instability, we will consider $z$ as an ignorable coordinate.\\
In this fluid model, the non-ideal term allowing magnetic field lines reconnection is associated with the finite electron mass. This model also allows for an anisotropic equilibrium electron temperature. The two independent parameters representing the above mentioned effects are 
\begin{equation}
      \delta^2= \frac{m_e}{m_i}, \quad   \Theta_e=\frac{T_{0_{\perp _e}}}{T_{0_{ \parallel _e}}},
\end{equation}
respectively.\\
Equation (\ref{continuity}) is obtained from the continuity equation for the electron gyrocenters and Eq. (\ref{ohmslaw}) results from the parallel component of the generalized Ohm's law.  These equations correspond to a particular limit of a more general gyrofluid model. Although such gyrofluid parent model is new, we postpone its full derivation in Appendix \ref{app:mod} in order not to overload the main text with the related technical details.\\
The model (\ref{continuity}) - (\ref{ohmslaw}) assumes the following ordering :
\beq \label{ordering2}
 \de^2 \ll \bpe  \ll 1.
\eeq
Considering a low $\bpe$ regime allows us to neglect the electron FLR effects in the parent model, while taking $\delta^2 \ll \bpe$ allows us to retain the terms proportional to electron inertia in Eq. (\ref{ohmslaw}) although  as small contributions. This is also related to the choice of $\rhos$ as characteristic length. Indeed, assuming $\de^2 \ll \bpe$ amounts to saying that the electron skin depth, i.e. the characteristic scale for inertial reconnection, is much smaller than the perpendicular sonic Larmor radius $\rhos$. This allows us to treat the non-ideal term, causing reconnection, as a small perturbation, and consequently to apply a perturbative approach.\\
It can be relevant to point out that, within the limit (\ref{ordering2}), the model admits the following static relations
\begin{equation} \label{ne}
    N_e = \lapp \phi, \qquad U_e  =  \Delta_{\perp} A_{\parallel}, \qquad \bpar = - \frac{\bpe}{2} \lapp \phi.
\end{equation}
These relations result from the quasi-neutrality condition and the perpendicular and parallel components of the Ampere's law respectively. 

Electron temperature anisotropy, on the other hand, is supposed to be weak, i.e. $\Theta_e $ is assumed to remain finite as $\delta$ and $\bpe$ tend to zero. Ions are assumed to be cold, with isotropic temperature, and all ion gyrocenter fluctuations are assumed to be negligible with respect to the electron ones, in the static relations. Consequently, ion gyrocenter fluctuations get decoupled from the system. 

When taking the isotropic limit $\Theta_e=1$ for the electron temperature, the system (\ref{continuity}) - (\ref{ohmslaw}) corresponds, up to the normalization, to a two-field model derived in Ref. \cite{Sch94}. However, as will be pointed out in Sec. 3,  unlike the present model, the  model of Ref. \cite{Sch94} has a characteristic length scale $L \gg \rhos$. As a consequence, in the latter model, the analogous of the last term on the right-hand side of Eq. (\ref{ohmslaw}), is typically a small perturbation, proportional to $(\rhos/L)\ll1$, whereas, on the presently adopted scale $\rhos$, this term is comparable to other terms retained in the same equation.  Neglecting electron inertia as well as the terms $-(1/\Theta_e)[\lapp \phi , \apar]$ and $(1/\Theta_e)\partial_z \lapp \phi$ in Eq.  (\ref{ohmslaw}), the system corresponds to low $\beta$ reduced magnetohydrodynamics \cite{Kad74,Str76}. A 2D version of a similar model, neglecting electron inertia but accounting for parallel magnetic perturbations, was recently adopted in order to investigate the impact of electron temperature anisotropy on the linear stability of magnetic vortex chains \cite{Gra20b}.

Using the relation $\bpar=- \frac{\bpe}{2} \lapp \phi$ (coming from Eq. (\ref{ne})) in the Eqs. (\ref{continuity}) and (\ref{ohmslaw}), the resulting model is similar to electron magnetohydrodynamic (EMHD) \cite{Kin90}. Analogously to EMHD, ions are assumed to be immobile along the guide field direction. On the other hand the continuity equation (\ref{continuity}) turns to be an evolution equation for the perturbation of the parallel magnetic field $\bpar$, as is the case for EMHD, and not of the field $\bpar - (2 \delta^2/ \bpe)\lapp \bpar$. Regarding the terms $-(1/\Theta_e)[\lapp \phi , \apar] + (1/\Theta_e)\partial_z \lapp \phi$, in Eq. (\ref{ohmslaw}), where the temperature anisotropy parameter $\Theta_e$, appears in the denominator, and is usually not present in EMHD, these are contributions from the projection, along the magnetic field, of the divergence of the anisotropic pressure tensor. The first of these two terms plays an important role in the linear stability analysis that we present in this paper. In particular, with respect to the case of EMHD, it leads to modifications of the linearized system in the inner region, eventually leading to a dispersion relation different from those derived for EMHD systems in Refs. \cite{Bul92, Cai08}.\\
In Refs. \cite{Kun95,Cai09} a fluid description for the electron species was adopted, in order to study collisionless tearing stability taking into account non-gyrotropic terms. In our stability analysis we do not account for such terms, as they are associated with electron FLR corrections, that we neglect due to the small $\bpe$ limit. Compared to such studies, where a space-dependent equilibrium pressure is also considered, our equations (\ref{continuity}) - (\ref{ohmslaw}) are assuming, in terms of the perpendicular and parallel particle temperature fluctuations, an isothermal closure and we consider a homogeneous equilibrium pressure resulting from a bi-Maxwellian distribution function.
This closure leads to the cancellation of a number of contributions coming from the pressure tensor and the terms $-(1/\Theta_e)[\lapp \phi , \apar] + (1/\Theta_e)\partial_z \lapp \phi$ are the remaining contributions.

It may be of interest to emphasize that the model (\ref{continuity}) - (\ref{ohmslaw}) possesses a noncanonical Hamiltonian structure \cite{Mor98} that can be derived following the procedure given in Ref. \cite{Tas19}. Its Hamiltonian is given by
\begin{equation}
    H(N_e, A_e)= \frac{1}{2} \int d^3 x \, \left( \frac{N_e^2}{\Theta_e} - A_e \lapp \lba A_e  - N_e \lapp^{-1} N_e \right),
    \label{eq:hamappendix}
\end{equation}
where $A_e = \apar - \frac{2 \de^2}{\bpe} \lapp \apar$ and  $\lba = \big(1 - \frac{2 \de^2}{\bpe} \lapp \big)^{-1} $ is a linear operator in terms of which one has $ \apar = \lba A_e$. The Poisson bracket is
\begin{equation}
\begin{split}
     & \{ F , G \}= \int d^3 x \, \Bigg( N_e \left(  [F_{N_e}, G_{N_e}]  + \frac{2 \delta^2}{\bpe \Theta_e}[F_{A_e} , G_{A_e}]\right)   \\
     & + A_e([F_{A_e} , G_{N_e}] + [F_{N_e} , G_{A_e}]) + F_{N_e} \frac{\partial G_{A_e}}{\partial z} + F_{A_e} \frac{\partial G_{N_e}}{\partial z}  \Bigg),
    \label{eq:pbappendix}
\end{split}
\end{equation}
acting on two functionals $F$ and $G$. In Eq. (\ref{eq:pbappendix}) the subscript on the functionals indicates functional derivative, so that, for instance $F_{N_e}=\delta F/\delta N_e$. The system can then be written in the Hamiltonian form : $\partial_t F = \{F, H\}$, where $F$ is an observable of the system such as, for instance, the vorticity $\lapp \phi$ or the parallel electron canonical momentum $A_e$. The adopted model (\ref{continuity}) - (\ref{ohmslaw}) therefore preserves the Hamiltonian character of the gyrokinetic model taken as starting point for its derivation \cite{Tas19}.

\section{Analytical investigation of the linear tearing mode stability}  \label{sec:lintear}

In this Section we consider the model equations linearized about an equilibrium state and derive a dispersion relation providing the growth rate of the linear tearing mode as function of various parameters of the system, among which the equilibrium electron temperature anisotropy.
For the equilibrium state we made the choice of a Harris sheet  with no background flow given by
\beq \label{equilibrium}
\apar^{(0)} (x)= - \lambda \ln \cosh (x/\lambda), \quad \phi^{(0)}(x) = 0,
\eeq
giving the equilibrium magnetic field with components $B_x^{(0)}=0$ and $B_y^{(0)}(x) = \sqrt{\bpe/2} \tanh(x /\lambda) $. The positive parameter $\lambda$ can be seen as a stretching factor for the equilibrium shear length given by $L_s=\lambda \rhos$, and defined as being the characteristic scale at which the dimensional reconnecting magnetic field is varying, so that $d_{\hx} \hat{B}_y^{(0)}(0) =  B_0 \sqrt{\bpe /2}/L_s$.
This equilibrium has no $y$-dependency to allow a Fourier analysis in this direction.
We linearize Eqs. (\ref{continuity}) and (\ref{ohmslaw}) considering perturbations of the form
\beq \label{perturbations}
\apar^{(1)} (x,y,t) = \frac{1}{2}(\wapar (x) e^{\gamma t +i k_y y} + \bar{\tilde{A}} (x) e^{\gamma t -i k_y y}) , \quad \phi^{(1)}(x,y,t) = \frac{1}{2}(\wphi(x) e^{\gamma t +i k_y y} + \bar{\tilde{\phi}}(x) e^{\gamma t -i k_y y}) .
\eeq
where $\gamma$ is the growth rate of the instability, $k_y = 2 \pi m/ L_y$ is the wave number, with $m \in \mathbb{N}$ and the overbar refers to the complex conjugate. This configuration has a resonant surface at $x=0$ and we selected perturbations independent on $z$ which effectively leads us to consider a $2D$ reduction of the model (\ref{continuity})-(\ref{ohmslaw}).  
The perturbations are subject to the boundary conditions $ \wapar , \,  \wphi \rightarrow 0$, as $x \rightarrow \pm  \infty$. In addition, we are seeking for even solutions of $\wapar(x)$ and odd solutions for $\wphi(x)$. This is a standard parity of the linear tearing mode solutions, which is respected by our model (\ref{continuity}) - (\ref{ohmslaw}). As a consequence of such parities, $\wapar$ and $\wphi$ are purely real and imaginary-valued functions, respectively.

Proceeding to the linearization of Eqs. (\ref{continuity}) and (\ref{ohmslaw}) gives
\begin{equation} \label{linearizedcont}
g (\wphi''  - k_{y}^{2} \wphi) -  i B_{y}^{(0)''} \wapar  + i B^{(0)}_{y} ( \wapar'' - k_{y}^{2} \wapar ) = 0, 
\end{equation}
\begin{equation} \label{linearizedohm}
\begin{split}
    g\Bigg( \wapar -  &  \frac{2  \de^2 }{\bpe} ( \wapar'' - k_{y}^{2} \wapar ) \Bigg) + i \wphi \left( \byo -    \frac{2  \de^2 }{\bpe}  B_{y}^{(0)''} \right) \\
    & -   \frac{i \byo}{\Theta_e} ( \wphi'' - k_{y}^{2} \wphi )  =0,
    \end{split}
\end{equation}
where the prime notation denotes the derivative with respect to the argument of the function and where
\beq
g=\frac{\gamma}{k_y}.
\eeq
We consider the time variation of the perturbation being slow, which, together with the assumption (\ref{ordering2}), leads to considering the following two small parameters
\beq \label{assumptions}
g \ll 1, \qquad \frac{2\de^2}{\bpe} \ll 1. 
\eeq
As is customary with linear tearing modes, the stability analysis leads to a boundary layer problem implying that one has to calculate the solution in two separate regions involving two different scalings. In the inner region, close to the resonant surface $x=0$, terms associated with electron inertia become significant and gradients of the perturbations are large. On the other hand, in the outer region, complementary to the inner region, terms proportional to electron inertia can be neglected. 
Once the two solutions are found in the two regions, they must be asymptotically matched.

\subsection{Outer region}   \label{ssec:outer}
In the limit where electron inertia is negligible, Eqs. (\ref{linearizedcont}) and (\ref{linearizedohm}) become  
\begin{equation}\label{outercont}
    g ( \wphi ''_{out} -  k_y ^2 \wphi_{out} ) -    i B_{y}^{(0)''} \wapar_{out}  +  i \byo  ( \wapar''_{out} - k_y ^2 \wapar_{out} ) = 0, 
\end{equation}
\begin{equation}  \label{outerohm}
\begin{split}
    g  \wapar_{out}  + i\byo \wphi_{out}  -   \frac{ i\byo}{\Theta_e}   ( \wphi''_{out} - k_y^2 \wphi_{out} ) =0.
    \end{split} 
\end{equation}
We then neglect the terms proportional to $g $ in Eq. (\ref{outercont}) and using  the equilibrium $\byo = \tanh (x /\lambda)$ we get the following equations
\beq \label{outereq1}
\wapar''_{out} - \wapar_{out}\left( k_y^2 - \frac{2}{\lambda^2 \cosh{(x/\lambda})^2} \right) =0,
\eeq
\beq \label{outereq2}
\wphi''_{out} - \left( k_y^2 + \Theta_e \right) \wphi_{out}  = - \frac{ i g \, \Theta_e}{ \byo} \wapar_{out}.
\eeq\\
We remark that, although Eq. (\ref{outereq1}) corresponds to a standard equation in MHD linear  tearing mode theory, yielding the outer solution for the perturbation $\widetilde{A}_{out}$ in the presence of a Harris sheet equilibrium, this is not the case for Eq. (\ref{outereq2}). In particular, the first two terms in Eq. (\ref{outereq2}) are absent in the outermost region in the tearing mode linear analysis of reduced MHD and of the two-field model of Refs. \cite{Caf98}. 
This difference comes from the fact that the structure of the reconnection region has a characteristic scale length corresponding to $\rhos$, in contrast with the current sheet of macroscopic length characteristic of MHD. The new terms in Eq. (\ref{outereq2}) are due to electron compressibility, which becomes relevant on the scale length $\rhos$.  They do not affect the outer solution for the magnetic flux function (and consequently the expression for the parameter $\Delta'$ defined later in Eq. (\ref{Delta'})), but alter the outer solutions for the flow.

The solution of Eq. (\ref{outereq1}) is given by: \cite{Whi86}
\begin{equation} \label{outersolutionA}
    \wapar_{out} (x)= \frac{e^{- k_y x}}{\lambda} \left( \frac{\tanh (x/\lambda)}{k_y}+\lambda\right),
\end{equation} 
whereas the solution of (\ref{outereq2}) can be found by the method of the variation of parameters and corresponds to
\beq \label{outersolutionphi}
\begin{split}
\wphi_{out} (x) = & i e^{x \alpha} \left(C_1 -\coeff  \int_{a}^{x} \left(  \frac{1}{k_y} + \frac{\lambda}{\tanh (t/\lambda)} \right) e^{- (\alpha + k_y) t} dt \right) \\
 & + i e^{- \alpha x} \left( C_2 + \coeff \int_{a}^{x} \left( \frac{1}{k_y} + \frac{\lambda}{\tanh (t/\lambda)} \right) e^{( \alpha - k_y) t }dt \right),
\end{split}
\eeq
where we introduced the  short-hand notation $\alpha^2 = k_y^2 + \Theta_e$. The lower integral bound $a$ is a strictly positive arbitrary value that can be freely chosen.  The constant $C_1$ is chosen to ensure that the boundary condition $ \lim_{x \rightarrow + \infty} \wphi_{out} = 0$ is respected and is given by
\beq
C_1 = \coeff   \int_{a}^{\infty} \left(  \frac{1}{k_y} + \frac{\lambda}{\tanh (t/\lambda)} \right) e^{- (\alpha + k_y) t} dt .
\eeq
Note that the coefficient $\alpha - k_y$, present in the exponential of the second term of $\wphi_{out}$, is positive. The convergence,  which might be not obvious at first sight, of this term in the limit $x \rightarrow \infty$, is shown in detail in Appendix \ref{app:lim}.\\
Although the solution $\wphi_{in}$ of the inner region is unknown at this stage, we anticipate in the following equation, the expression of the constant $C_2$, which has been chosen based on the matching condition between inner and outer solution for $\wphi$:
\beq
C_2 =  g \Theta_e \lambda a -  g^2 \de \sqrt{\frac{2}{\bpe}} \lambda^2 \Theta_e^{3/2} \frac{\pi}{2} - C_1.
\eeq
 The derivative of the outer solution (\ref{outersolutionA}), $ \wapar_{out} '$, is  discontinuous at the resonant surface and the discontinuity is usually measured with the standard tearing parameter
\begin{equation} \label{Delta'}
    \Delta' =  \lim_{x \rightarrow 0^{+}} \frac{\wapar_{out}'}{\wapar_{out}}  - \lim_{x \rightarrow 0^{-}} \frac{ \wapar_{out}'}{\wapar_{out}}  ,
\end{equation}
with $\Delta' >0$ implying instability for the standard MHD case with static equilibrium. 
Computing $\Delta'$ for the solution (\ref{outersolutionA}) gives
\beq \label{Delta'outer}
\Delta' = \frac{2}{\lambda}\left( \frac{1}{k_y \lambda} - k_y \lambda\right). 
\eeq
In the limit $|x| \rightarrow 0 $ the solution can then be written in the form 
\beq \label{expansionA}
\wapar_{out}=  1 + \frac{\Delta'}{2} |x| + O(x^2).
\eeq
In this work we consider small values of $\Delta'$ allowing the use of the so-called \textit{constant psi approximation} which consists in approximating $\wapar$ as a constant close to $x=0$ \cite{Fur63}.  
\subsection{Inner region} \label{ssec:inner}

In the inner region, centered about the resonant surface, where $x\ll1$, we introduce the change of variable 
\beq \label{changeofvariable}
x =\epsilon \bar{x}, \quad \mbox{with} \, \, \,  \epsilon \ll 1,
\eeq
and consider the unknown functions $\wapar_{in}(\bar{x})$ and $\wphi_{in} (\bar{x})$ defined by $\wapar (x)=\wapar_{in}(\bar{x})$ and $\wphi (x)=\wphi_{in} (\bar{x})$.

Assuming that $\epsilon$ is a small parameter implies that x-derivatives are large in this region  (i.e.  $k_y\ll \partial_x $) since they will scale as $1/\epsilon$. Moreover, $\byo(x)=\byo(\epsilon \bar{x})$ can be substituted by its Taylor expansion  $\epsilon \rightarrow 0$. Inspection of the ordering of the various terms of Eqs. (\ref{linearizedcont}) - (\ref{linearizedohm}) in the inner region, after rescaling, indicates that
\beq \label{epsilon}
\epsilon = g \delta \sqrt{\frac{2}{\bpe}},
\eeq 
is the distinguished limit allowing to keep the maximum number of terms in the system as $\epsilon \rightarrow 0$ \cite{Ben99}. Thus (\ref{epsilon}) provides the appropriate choice for the scaling parameter $\epsilon$.  

The leading contributions of Eqs. (\ref{linearizedcont}) and (\ref{linearizedohm}) in the inner region are
\beq \label{innercont}
\wapar_{in}'' = \frac{i g\lambda }{\bar{x} \epsilon}  \wphi_{in}'',
\eeq
\beq 
\label{innerohm}
g \left( \wapar_{in}  -  \frac{2 \de^2}{\bpe} \frac{1}{ \epsilon} \wapar_{in}'' \right)-  \frac{i \bar{x}}{\Theta_e \lambda}\frac{1}{ \epsilon}  \wphi_{in}'' = 0 .
\eeq
Inserting (\ref{innercont}) into (\ref{innerohm}) we get 
\beq \label{phi''}
\frac{i }{\epsilon} \wphi_{in}'' = \frac{ g \bar{x} \wapar_{in}}{\lambda( 1 +  \frac{\bar{x}^2}{\Theta_e \lambda^2})} . 
\eeq
We introduce a re-scaled displacement function $\hxi$ related to $\wphi_{in}$ by 
\beq \label{dispfunction}
\hxi = - \frac{i }{g\epsilon} \wphi_{in}.
\eeq
The rescaling in Eq. (\ref{dispfunction}) allows to eliminate the  inner parameter $\epsilon$ from Eq. (\ref{phi''}) and gives the following layer equation for $\hxi$
\beq \label{layer}
 \hxi''=\frac{- \bar{x}\wapar_{in} }{\lambda  \left( 1+  \frac{ \bar{x}^2}{\Theta_e \lambda^2} \right)}.
\eeq
Under the change of variables (\ref{changeofvariable}), the expansion (\ref{expansionA}) becomes $\wapar_{out}=  1 + \frac{\Delta'}{2} \epsilon |\bar{x}| + O(\bar{x}^2).$ 
As stated before, we consider small values of $\Delta'$ so we can apply the constant-$\psi$ approximation.
Thus, we can set $\wapar_{in} = 1 +\wapar_1(\bar{x})$,
where $\wapar_1(\bar{x}) \ll 1$.
Combining Eqs. (\ref{innercont}) and (\ref{layer}) one eventually finds  that the solution for $\hxi$ is not required for determining the dispersion relation (which will be obtained by calculating the integral in Eq. (\ref{intA''})). Nevertheless, the solution of Eq. (\ref{layer}) has to be determined to make sure the matching does exist in the overlap region. Such solution is given by
\beq
\hxi(\bar{x}) = - \frac{\lambda \Theta_e}{2}  \bar{x} \log \left(\lambda ^2\Theta_e +\bar{x}^2\right) - \lambda^2 \Theta_e^{3/2}  \arctan\left(\frac{\bar{x} }{\lambda \Theta_e^{1/2}}\right)  +\bar{x}( \lambda \Theta_e  + D_2 ) + D_1.
\eeq
We set $D_1=0$ in order to respect the condition $\lim_{\bar{x} \rightarrow 0} \hxi =0$, following from $\wphi$ being an odd function.\\
In terms of electrostatic potential $\widetilde{\phi}_{in}$ and of the outer variable, $\hxi(\bar{x})$ reads
\beq
\begin{split}
\wphi_{in} (\epsilon x) =  & - \frac{i\Theta_e \lambda g }{2} x  \log\left(\lambda^2 \Theta_e  + \frac{x^2 \bpe}{g^2 \de^2 2}  \right)   + i g x(\lambda \Theta_e  + D_2) \\
& - i g^2 \de \sqrt{\frac{2}{\bpe}} \lambda^2 \Theta_e^{3/2} \arctan\left( \sqrt{\frac{\bpe}{2}} \frac{x}{ g \de\lambda \Theta_e^{1/2}}\right),
\end{split}
\eeq
where $D_2$ has been chosen as
\beq
D_2 = \frac{\alpha}{ g}( C_1 - C_2) + \frac{a \Theta_e}{k_y \lambda} - a k_y \lambda \Theta_e + \lambda \Theta_e \log\left( \sqrt{\frac{\bpe}{2}}\frac{a}{g \de}\right). 
\eeq
The choice we made for $C_2$ and $D_2$ ensures that $\wphi_{out}$ and $\wphi_{in}$ have the same expansion in the overlap region. \\
As customary in linear tearing mode theory, we add the following matching condition concerning the derivatives of the solutions
\begin{equation} \label{intA''}
    \Delta'= \frac{1}{\epsilon} \int^{\infty}_{-\infty} \wapar''_{in} d\bar{x}.
\end{equation}
This condition (\ref{intA''}) will provide the dispersion relation giving the growth rate of the tearing mode as function of $\Delta'$ and of other parameters of the system.
Combining  Eqs. (\ref{innercont}), (\ref{dispfunction}) and (\ref{layer}) with the relation (\ref{intA''}), and making use of the constant-$\psi$ approximation, leads to 
\beq
\Delta'= \frac{2 g^2}{\epsilon} \int^{+ \infty}_{0}\frac{d \bar{x}}{\left( 1 + \frac{\bar{x}^2 }{\Theta_e \lambda^2}\right)}.
\eeq
Using the expression (\ref{epsilon})  for $\epsilon$, we obtain the following dispersion relation
\beq \label{growthrate}
\gamma = \frac{ \Delta' k_y \de }{\pi \lambda \sqrt{\Theta_e}}\sqrt{\frac{2}{\bpe}},
\eeq
where for $\Delta '$ one has to use the expression (\ref{Delta'outer}).

The dispersion relation (\ref{growthrate})  accounts for well known features of inertial reconnection in the small $\Delta '$ regime, such as the linear dependence on $\Delta '$ and on the square root of the mass ratio $\delta$ \cite{Por91, Fit07,Fit10}. The growth rate (\ref{growthrate}), for $\Theta_e=1$,  reduces to the one found in Ref. \cite{Tas18}, the difference being the factor $\sqrt{2/ \bpe}$ emerging from the fact that we used a different normalization of the perturbed magnetic potential, which amounts to taking a different dimensional magnetic equilibrium. The main element of novelty is given by the dependence on the electron temperature anisotropy parameter $\Theta_e$. The formula (\ref{growthrate}) predicts a decrease of the reconnection growth rate as $\Theta_e$ increases. This differs from what occurs in the case of weak or absent guide field \cite{For68,Che84,Shi87,Kar04,Que10} where, as already mentioned in Sec. \ref{sec:intro}, an enhancement of the growth rate is observed as the temperature anisotropy parameter increases.

In order to make the dependence of the growth rate (\ref{growthrate}) on various physical quantities, more transparent, and to facilitate the comparison with the formula derived in Ref. \cite{Por91} for the isotropic case, we rewrite the relation (\ref{growthrate}) in the following dimensional form: 
\beq \label{dimgrowth}
\hat{\gamma}= \frac{2}{\lambda \rhos}\left( \frac{1}{\hat{k}_y\lambda\rhos}-\hat{k}_y\lambda\rhos\right)\frac{ \hat{k}_y v_{th_{\parallel_e}} d_e^2 }{\pi \lambda d_i} ,
\eeq

where we used the dimensional expression for the tearing parameter $\hat{\Delta}'= \frac{2}{\lambda \rhos}\left( \frac{1}{\hat{k}_y\lambda\rhos}-\hat{k}_y\lambda\rhos\right)$,
which allows to see its dependence in $\rhos$, and consequently, its dependence in $\Tpee$ (we recall that $\rhos \propto \sqrt{\Tpee}$). In the dimensional expression, the dependence in $\Tpee$ coming from the parameters $\bpe$ and $ \Theta_e$ of the normalized expression (\ref{growthrate}), is canceled. This implies that the equilibrium perpendicular temperature appears in the final expression (\ref{dimgrowth}) from the fact that it affects the current sheet width, and consequently the tearing parameter as well. In Eq.  (\ref{dimgrowth}), we used the expression for the electron thermal speed based on the parallel electron temperature $v_{th_{\parallel_e}} =\sqrt{\Tpae/m_e}$. The parameters $d_i$ and $d_e$ are the ion and electron skin depths respectively, defined as $d_r=\sqrt{c^2 m_r/e^2 n_0 4 \pi}$, for $r=e,i$. \\
We point out that, according to our normalization (\ref{norm}) and to our choice (\ref{equilibrium}) for the equilibrium magnetic flux function, we are considering a dimensional equilibrium magnetic field $\hat{B}_y^{(0)}(\hx) = B_0 \sqrt{\bpe/2} \tanh(\hx /(\lambda \rhos))$ whose amplitude contains the factor $\sqrt{\bpe/2}$ and is given by $B_0 \sqrt{\bpe/2}$. 
Considering this equilibrium, in the isotropic case, $\Theta_e=1$, and with $\lambda=1$, the dimensional growth rate is given by   \cite{Por91}
\beq  \label{gammaiso}
\hat{\gamma}_{iso}= \frac{\hat{\Delta}' \hat{k}_y v_{th_{_e}} d_e^2}{\pi L} \sqrt{\frac{\beta_e}{2}},
\eeq
where $L$ is the shear length, $v_{th_{_e}}=\sqrt{T_{0_e}/m_e}$ is the isotropic electron thermal speed and $\beta_e$ is the isotropic 
electron beta parameter. Our growth rate (\ref{growthrate}) reduces to the one given in Eq. (\ref{gammaiso}) in the absence of temperature anisotropy and for $L=\lambda \rhos$. Therefore, the comparison shows that the extension of the formula for $\hat{\gamma}_{iso}$, to account for equilibrium electron temperature anisotropy is obtained by replacing the electron thermal speed with the parallel electron thermal speed and by considering the appropriate equilibrium scale length. 
Note, on the other hand, that the dispersion relation (\ref{dimgrowth}) differs from those derived in Refs. \cite{Bul92, Cai08} for EMHD models. In particular, for such models, the growth rate depends quadratically on $\Delta'$, whereas, in our case, the dependence is linear.

\section{Numerical results}   \label{sec:num}
\subsection{Validation of the dispersion relation}

\begin{figure}[!ht]
    \centering
\includegraphics[trim=16 0 0 0, scale=0.9]{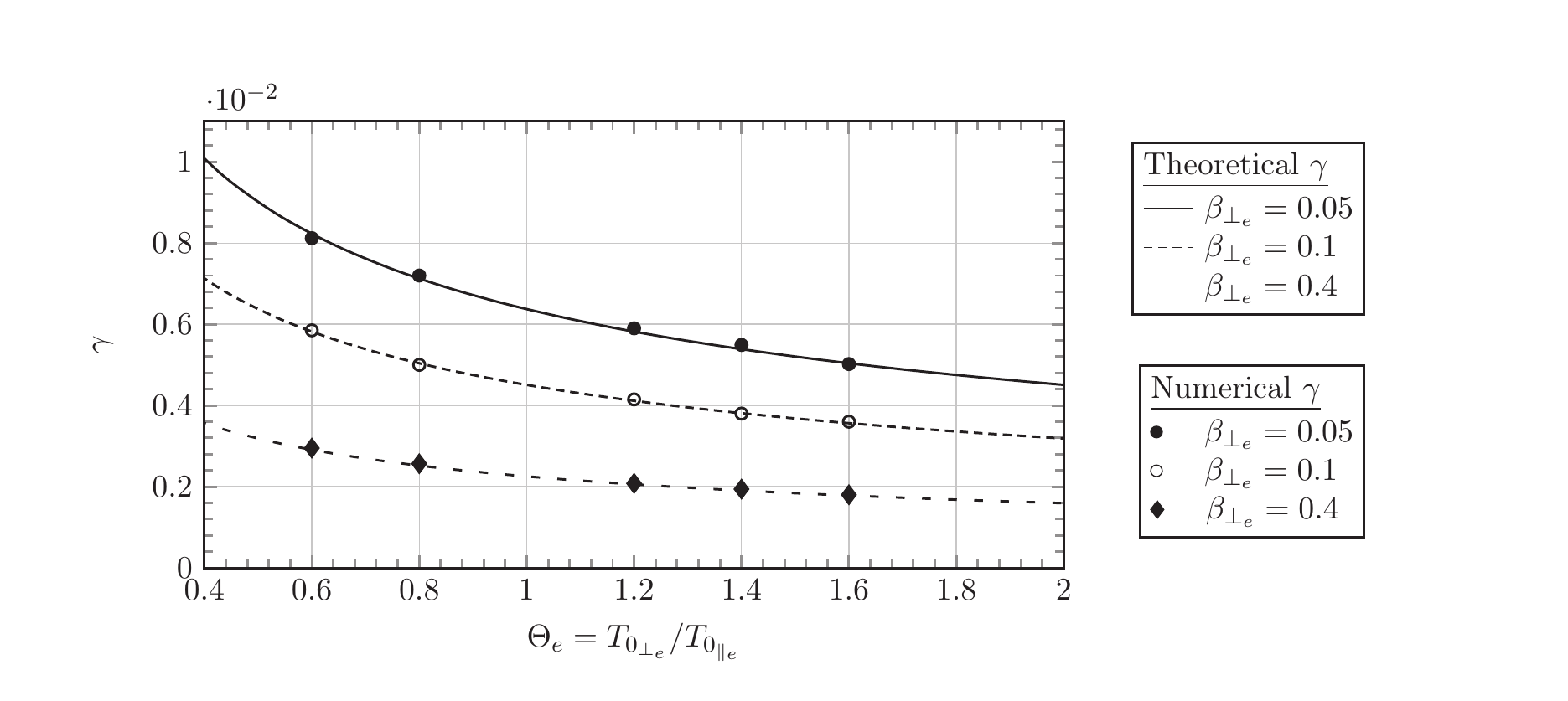}
    \caption{ \normalsize  Growth rate as a function of the electron temperature anisotropy for different values of $\bpe$ and taking the parameters $\delta = 0.1$ and $\Delta'=0.38$ fixed. For the full circle, the run has been made taking $n_x= 300$, for the empty circles $n_x=600$ and for the diamonds $n_x=1200$.  We recall that $\gamma$ is normalized with respect to $ \omega_{ci}$, the ion gyrofrequency.}
    \label{fig:gtg}
\end{figure}

Simulations have been carried out in order to test the analytical dispersion relation (\ref{growthrate}) with the code used in Refs. \cite{Gra20,Tas18}, which has been adapted to the new set of equations. The advancement in time is achieved through a third order Adams-Bashforth scheme. Periodic boundary conditions are imposed along the $x$ and the $y$-directions. A resolution of $n_y=160$ has been adopted and $n_x$ as been adapted according to the small parameter $2 \delta^2/\bpe$.
Although the code is available in 3D, we made the use, for this study, of a 2D domain given by $\{- L_x \leq x \leq L_x, -L_y \leq y \leq L_y \}$, where $L_y=4 \pi $ and $L_x = 10 \pi$. We made the choice of $\lambda=3$ which, according to the relation (\ref{Delta'outer}), gives the fixed values $\Delta' = 0.38$ with $k_y=0.24$ for the mode $m=1$. The mass ratio has been taken equal to $\delta^2=0.01$ while different values of $\bpe$ have been considered.  We point out that, considering that $\bpe = 2 \rhos^2/d_i^2$, where $\rhos$ can be explicitly written as $\rhos= \sqrt{\Tpee m_i c^2 / e^2 B_0^2}$, a variation of $\bpe$ can be interpreted in two different ways. On one hand, it can mean a variation of $d_i$ performed by modifying the background density $n_0$ and keeping a current sheet of a fixed thickness. Alternatively, it can also be interpreted as a variation of the perpendicular equilibrium temperature $\Tpee$, which implies a variation of the thickness of the current sheet $\rhos$, for a fixed density $n_0$. Both ways leave $B_0$ and $m_i$ (and in turn $\omega_{ci}$) constant, thus guaranteeing that one is comparing different values of the growth rate, normalized with respect to the same unit of time. \\
The numerical growth rate is determined by the formula 
\beq \label{numericalgrowthrate}
\gamma_N = \frac{d}{dt} \log\left| \apar^{(1)} \left(\frac{\pi}{2},0,t \right)  \right|,
\eeq
evaluated during the linear phase. 
In the definition (\ref{numericalgrowthrate}), the perturbed magnetic flux is evaluated at $(x=\pi/2, y=0)$ corresponding to the magnetic island $X$-point. 
Figure \ref{fig:gtg} shows a very good agreement between the analytical formula (\ref{growthrate}) and the numerical results. In particular, a stabilizing, although weak, role of electron temperature anisotropy at a fixed $\bpe$ is clearly visible, as $\Theta_e$ increases. The stabilizing role, for the adopted normalization, of the $\bpe$ parameter is also confirmed.
%
%
%
%

\subsection{Limits of validity of the dispersion relation}
The dispersion relation (\ref{growthrate}) was derived on the basis of a number of assumptions imposing some parameters to be small. We find it useful to summarize and discuss here such assumptions.
The adopted model (\ref{continuity}) - (\ref{ohmslaw}) neglects electron FLR effects. It is thus valid for scales much larger than the electron Larmor radius. This allowed in particular for a simplification, based on the relation $\delta^2 \lapp \ll 1$, in the form of the electron gyroaverage operators (\ref{approxG}). In particular, in the inner region, electron FLR corrections would lead to the presence of operators of the form $(\delta^2 / \epsilon^2) (\partial^2 / \partial \bar{x}^2)$. Therefore, in order for our linear analysis to be consistent with the assumptions of the model, the order of the inner scale, $\epsilon$, has to satisfy the condition
\beq \label{condflr}
\de \ll \epsilon.
\eeq
The relation (\ref{condflr}) and the other assumptions adopted for the tearing analysis are listed in Table 1. Given the expressions for $\epsilon$ (\ref{epsilon}) and for $\gamma$ (\ref{growthrate}),  it is possible to see that the assumptions 1, 2 and 3 are compatible.  Combining the assumptions 4 and 5 gives the following additional constraint on $\Delta'$
\begin{equation} \label{rangeDelta'}
\frac{\pi \lambda \sqrt{\Theta_e}}{\de}\frac{\bpe}{2} \ll \Delta' \ll \frac{\sqrt{\pi \lambda \sqrt{\Theta_e}}}{\de}\sqrt{\frac{\bpe}{2}}.
\end{equation}
The assumption  5, for neglecting the electron FLR effects in the inner region, provides the lower bound of (\ref{rangeDelta'}). It compels us, in order to be perfectly consistent with the derivation of the model, not to take too small values for $\Delta'$. Fulfilling such condition forces the admissible values for $\Delta '$ to lie in a rather narrow interval, given in Eq. (\ref{rangeDelta'}). Although the derivation of the dispersion relation (\ref{growthrate}), starting from the system (\ref{continuity}) - (\ref{ohmslaw}), was obtained using only  assumptions 1 - 4, violating the assumption 5 leaves the doubt about whether electron FLR effects might have played an important role in the inner region. 
\begin{table}
\caption{Table summarizing the various assumptions identifying small parameters adopted to derive the relation (\ref{growthrate}).}{\label{tab:table1}} 
\begin{tabular}{lllll}
 \hline\hline \\ \vspace{0.1cm}
 No. &  Adopted assumptions \\  \hline \\    \vspace{0.2cm}
 1& Time variation of the perturbation is slow   &  $g\ll 1$ \\  \vspace{0.2cm} 
 2& Smallness of the inner scale   &  $\epsilon \ll 1$   \\    \vspace{0.2cm}
 3& Keeping electron inertia terms while neglecting electron FLR  &  $\de^2 \ll \bpe \ll 1$   \\ \vspace{0.2cm}
 4& Use of the constant-$\psi$ approximation   &  $\epsilon \Delta' \ll 1$      \\\vspace{0.2cm}
 5& Neglecting electron FLR effects in the inner region &  $\de \ll \epsilon$   \\   
 \hline\hline \vspace{0.1cm}
\end{tabular}
\end{table}
In assumption 3, the condition $\bpe \ll 1$ allows to neglect the FLR effects in the general model while retaining electron inertia. Recalling that $\bpe = 8 \pi n_0 \Tpee / B_0^2$, the low-$\bpe$ regime is consistent with the strong guide field configuration.
 Although negligible at the leading order, if we keep first order corrections in $\bpe$ in the evolution equations, as they are written in Eqs.  (\ref{continuitywithbpe}) and (\ref{ohmslawwithbpe}), we find the modified growth rate
\beq \label{grlargebeta}
\gamma = \frac{\de \Delta' k_y}{\pi \lambda}\left( \frac{2}{\bpe} + 1 - \frac{1}{\Theta_e} \right) \left( \frac{1}{\Theta_e} - \frac{\bpe}{\bpe + 2} \right)^{1/2} \sqrt{\frac{\bpe}{2}} ,
\eeq
which is non-negative if
\beq  \label{boundtheta}
 \frac{\bpe}{2 + \bpe} \leq \Theta_e \leq  \frac{2 + \bpe }{\bpe}.
\eeq
By comparing Eq. (\ref{grlargebeta}) with Eq. (\ref{growthrate}), it emerges that the inclusion of first order corrections in $\bpe$ introduces, in the dispersion relation, new modifications due to the temperature anisotropy, which are not due to the specific choice of the equilibrium scale length. We remark that the condition on the lower bound for $\Theta_e$ in Eq. (\ref{boundtheta}), corresponds to the condition for firehose stability \cite{Has75}. However, under the adopted assumptions $\bpe \ll 1$ and $\Theta_e =O(1)$, violating this condition leads outside the range of validity of the model.
\begin{figure}[!ht]
    \centering
\includegraphics[trim=16 70 0 50, scale=1]{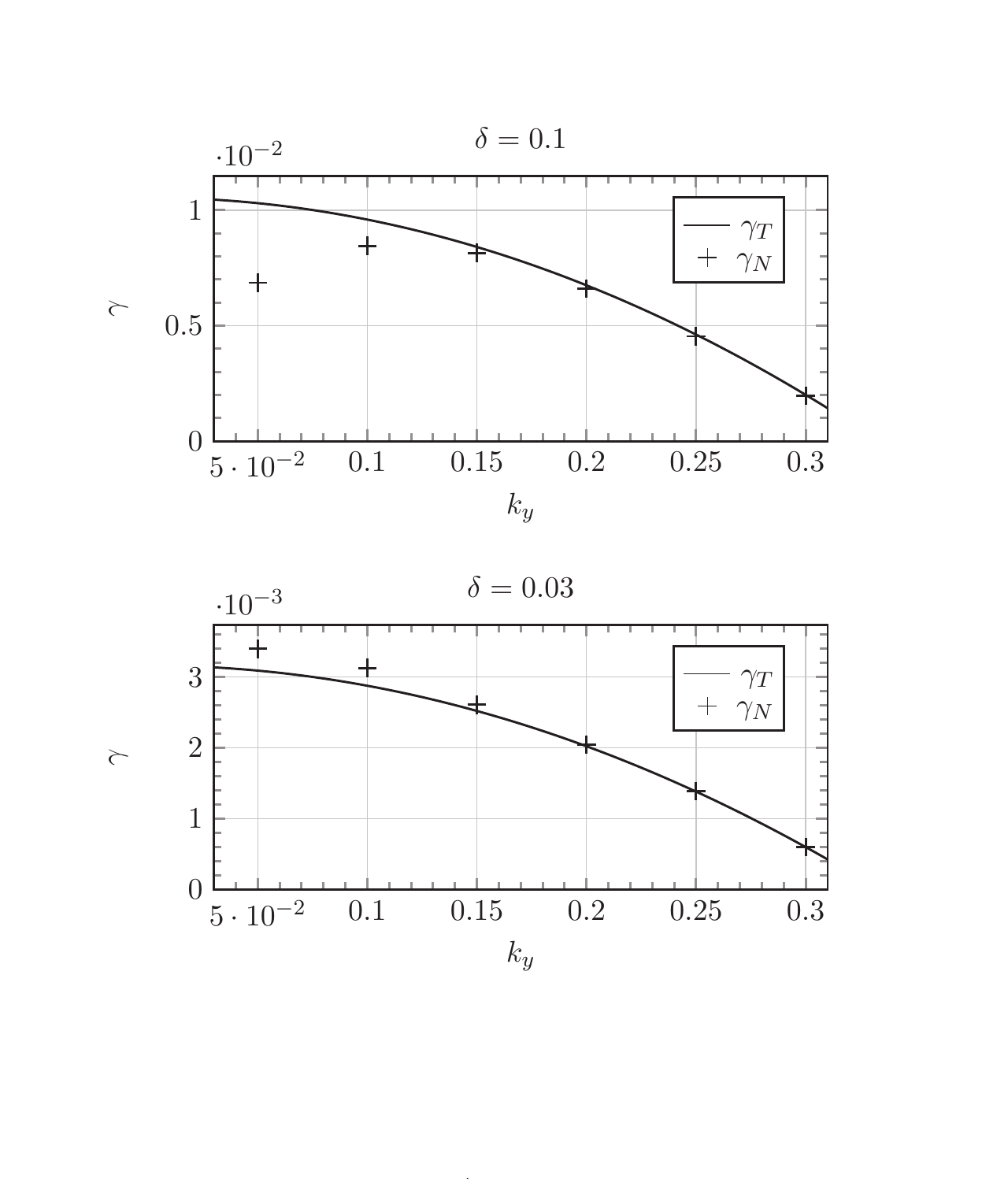}
    \caption{ \normalsize  Comparison between the theoretical growth rate predicted by Eq. (\ref{growthrate}) ($\gamma_T$) and the numerical growth rate ($\gamma_N$) as a function of the wave number $k_y$, for the cases $\delta=0.1$ and $\delta=0.03$. The values of the parameters are $\Theta_e=1$, $\bpe=0.1$, $\lambda=3$ and the numerical simulations were carried out using the modes $1 \leq m \leq 6$.}
    \label{fig:gky}
\end{figure}
We find it also useful to analyze how the theoretical predictions based on Eq. (\ref{growthrate}) compare with the numerical results, for different values of $k_y$. Figure 2 shows the numerical and theoretical growth rates for different $k_y$ values in the cases $\delta=0.1$ and $\delta=0.03$. We recall that the relation between $\Delta'$ and $k_y$  is given by Eq. (\ref{Delta'outer}). In order to obtain higher values of $\Delta'$ for low modes, the length $L_y$ has been increased to $L_y=20 \pi$, while we kept $L_x=10 \pi$. 
As shown in Fig. 2, in the two cases, the agreement between the theoretical and the numerical values is very satisfactory for $k_y > 0.15$ (corresponding to $\Delta' < 1.15$). 
For $\delta=0.1$, the lowest mode (associated to $k_y=0.05$, $\Delta'=4.3$) gives a relative error between $\gamma_T$ and $\gamma_N$ of $33\%$. In this case we have $\epsilon \Delta' = 0.4$ and $\gamma/k_y = 0.2$, which appear to be too large for the formula (\ref{growthrate}) to be valid. 
However, in the case of $\delta=0.03$, which leads to a smaller $\gamma/k_y$ ratio, the error for the lowest mode is reduced to $9\%$. In the latter case we have $\epsilon \Delta' =0.03$ and $\gamma/k_y = 0.06$. 
To conclude, the error between $\gamma_N$ and $\gamma_T$ is well decreasing with the ratio $\gamma/k_y$ and the product $\epsilon \Delta'$, in consistency with the assumptions No. 1 and 4 of Table 1 (note that $\epsilon$ depends linearly on $\gamma/k_y$). Small values of $ \gamma / k_y $ increase the upper bound of (44) and allow for larger values of $ \Delta'$.


\section{Conclusions}  \label{sec:concl}  

In conclusion, we showed that, in the strong guide field regime, the effect of equilibrium electron temperature anisotropy is different from what occurs in the case of absent or weak guide field. The enhancement of the growth rate observed in the latter case is indeed absent in the strong guide field case. In this respect we remark that, in Ref. \cite{Shi87}, the tendency of a finite guide field   to weaken such enhancement was observed (although with a guide field of amplitude at maximum only 2.5 greater than that of the equilibrium field). For a magnetic equilibrium shear length of the order of $\rhos$, as used in the present paper, we showed that the growth rate actually exhibits a weak decrease as $\Theta_e$ increases.
The various assumptions adopted in order to simplify the original gyrofluid model (\ref{nsnorm})-(\ref{tparnorm})  made the problem of tearing stability treatable analytically but, on the other hand,  impose severe restrictions on the range of applicability of the results. In the following, we list what we consider could be priorities for future work, allowing for more realistic applications.
In terms of modelling space plasmas, the cold ion assumption, for instance, is questionable and, in general, a more refined treatment of ion dynamics would be desirable. Accounting for electron FLR effects provides a further direction for improvement, whose investigation is currently in progress. In particular, relaxing the assumption $\de^2 \ll \bpe \ll 1$  would allow to study also the interaction of tearing modes with instabilities induced by temperature anisotropy. The inclusion of anisotropic temperature fluctuations in our general gyrofluid model (although this might cause some concerns with regard to the Hamiltonian structure, as pointed out in Ref. \cite{Tas16b}) could in principle make it possible to study numerically, in a gyrofluid context, fast reconnection induced by electron temperature anisotropy \cite{Cas15}. Finally, we also think that an investigation (to be carried out with a kinetic model) of the reconnection rate in the transition between the weak and strong guide field regimes, in the presence of temperature anisotropy, would help reconciling our present result with previous results not assuming a strong guide field. In the case of an equilibrium with isotropic temperature, this problem was treated in Ref. \cite{Dau05}.

\section*{Acknowledgments}

The authors wish to thank William Daughton for the helpful comments. 

\begin{appendix}

\section{Model derivation }   \label{app:mod}

In this Appendix we first derive a new gyrofluid model (corresponding to Eqs. (\ref{nsnorm})-(\ref{tparnorm})) accounting in particular for equilibrium temperature anisotropies, finite $\beta$ effects and isothermal closures for the {\it particle} temperature fluctuations. Subsequently we show how, from such model, the model (\ref{continuity})-(\ref{ohmslaw}), adopted in the tearing stability analysis, can be derived.

 We consider, as departure point, the gyrokinetic equations of  Ref. \cite{kunz2015}, although not in their most general form, as we specialize to the case of bi-Maxwellian equilibrium distribution functions and assume no equilibrium drift velocities. The reason for the choice of this parent gyrokinetic model is due to the fact that such model combines the relative simplicity of the reduced $\delta f$ approach with the inclusion of finite $\beta$ effects and of equilibrium temperature anisotropy, the latter in particular being  crucial for our tearing stability analysis.

The parent gyrokinetic model equations we consider are given by
\begin{align} \label{1}
 \frac{\partial \hga}{\partial \hht} & +\frac{c}{B_0}\left[ J_{0s} \hphi - \frac{\hvpar}{c} J_{0s} \hap +2 \frac{\hmua B_0}{\qa}J_{1s}\frac{\hbpar}{B_0} , \hga \right]
 \nno \\
& +\hvpar \frac{\partial }{\partial \hz}\left( \hga  +\frac{\qa}{\Tpas} \hcalfa\left( J_{0s} \hphi - \frac{\hvpar}{c} J_{0s}\hap +2 \frac{\hmua B_0}{\qa} J_{1s}\frac{\hbpar}{B_0}\right)\right)=0,
\end{align}
\begin{align} \label{2}
\sum_{s} \qa & \int \dwa \, J_{0s} \hga = \sum_{s} \frac{\qa^2 }{\Tpes}  \int \dwa \,  \hcalfa \left( 1 - J_{0s}^2 \right)  \hphi \quad   \nno \\
&- \sum_s \qa \int \dwa \, 2 \frac{\hmua B_0}{\Tpes} \hcalfa J_{0s} J_{1s} \frac{\hbpar}{B_0},  
\end{align}
\begin{align} \label{3}
\sum_s \qa & \int   \dwa \,  \hvpar J_{0s} \left( \hga -\frac{\qa}{\Tpas} \frac{\hvpar}{c} \hcalfa J_{0s} \hap\right) \nno \\
&= -\frac{c}{4 \pi} \lapp \hap + \sum_s \frac{\qa^2}{m_s}\int \dwa \,\hcalfa \left( 1 - \frac{1}{\thea}\frac{\hvpar^2}{\vtpa^2}\right)(1 - J_{0s}^2 ) \frac{\hap}{c},  
\end{align}
\begin{align} \label{4}
\sum_s  \frac{\bepes}{n_0} & \int  \dwa \, 2 \frac{\hmua B_0}{\Tpes} J_{1s} \hga= - \sum_s \frac{\bepes}{n_0} \frac{\qa}{\Tpes} \int \dwa \,  2 \frac{\hmua B_0}{\Tpes} \hcalfa J_{0s} J_{1s}  \hphi  \nno \\
&-\left(2 + \sum_s \frac{\bepes}{n_0}\int \dwa \, \hcalfa \left( 2 \frac{\hmua B_0}{\Tpes} J_{1s}\right)^2 \right) \frac{\hbpar}{B_0}. 
\end{align}
Consistently with the notation already adopted in Sec. \ref{sec:intro}, we indicated with a hat the dimensional dependent and independent variables. We also recall that the index $s$ denotes the particle species.
Equation (\ref{1}) describes the evolution of the generalized perturbed distribution function $\hga(\hx,\hy,\hz,\hvpar,\hmua ,\hht)$, which is connected to the perturbed gyrocenter distribution function $\hdfa$ by the relation
\beq  \label{pertdf}
\hga(\hx,\hy,\hz,\hvpar,\hmua ,\hht)=\hdfa(\hx,\hy,\hz,\hvpar,\hmua,\hht)+\frac{\qa}{\Tpas}\frac{\hvpar}{c}\hcalfa (\hvpar , \hmua) J_{0s} \hap (\hx,\hy,\hz,\hht),
\eeq
where $\hvpa$ is the velocity parallel to the equilibrium magnetic field and is used as a first velocity coordinate in phase space. Indicating with $\hvpe$ the perpendicular velocity, we express with $\hat{\mu}_s = m_s \hvpe^2 / 2 B_0$, the magnetic moment, used as a second velocity coordinate in phase space.
These two coordinates, together with the gyroangle $\theta$, allow us to define the integration over the velocity volume element,
\beq
 \int \dwa=  \int^{+ \infty}_{- \infty} d \hvpar \int^{+ \infty}_{0} \frac{2 \pi B_0}{m_s} d \hmua .
\eeq
Equations (\ref{2}), (\ref{3}), (\ref{4}) represent the quasineutrality
constraint and the parallel and perpendicular components of the Amp\`ere's law, respectively. We define by $\vtpa$ the parallel thermal speed, related to $\Tpas$ by $\vtpa = \sqrt{ \Tpas / m_s}$ and $q_s$ the charge. 
The equilibrium distribution function is a bi-Maxwellian whose expression is given by
\beq  \label{bimax}
\hcalfa(\hvpar,\hmua)= \left(\frac{m_{s}}{{2 \pi}}\right)^{3/2} \frac{n_0}{\Tpas^{1/2} \Tpes}\mathrm{e}^{-\frac{m_{s} \hvpar^2}{2 \Tpas}-\frac{\hmua B_0}{ \Tpes }},
\eeq
giving the uniform equilibrium density $n_0 = \int \dwa \hcalfa $. The equations (\ref{1}) - (\ref{4}) are valid for small perturbation of the equilibrium distribution function (\ref{bimax}). \\
The gyroaverage operators $J_{0s}$ and $J_{1s}$
are defined, in the Fourier space, as multiplications by $J_0(\aal)$ and $J_1(\aal)$, the latter being the zero and first order Bessel functions of the first kind, respectively. Therefore one has
\begin{align}
& J_{0s} f (\hx,\hy,\hz )=\sum_{\hbk } J_0 (\aal) f_{\hbk} e^{i \hbk \cdot \hbr},\\
& J_{1s} f (\hx,\hy,\hz)=\sum_{\hbk} \frac{J_1 (\aal)}{\aal} f_{\hbk} e^{i \hbk \cdot \hbr},
\end{align}
where the vector $\hbk = (\hat{k}_x,\hat{k}_y,\hat{k}_z) \in \{2\pi  l / (2 L_x \rhos), 2\pi m / (2 L_y \rhos), 2\pi n / (2 L_z \rhos) : ( l, m, n) \in \mathbb{Z}^{3}\}$ and  $\hbr \in \{(\hx,\hy,\hz) \in \mathbb{R}^3 | - \rhos L_x \leq \hx \leq \rhos L_x, - \rhos L_y \leq \hy \leq \rhos L_y, - \sqrt{2 / \bpe} \rhos L_z \leq \hz \leq \sqrt{2 / \bpe} \rhos L_z\}$. The quantity $a_s$ is defined by $a_s = \hat{k}_\perp \vpe / \omega_{cs}$ with $\hat{k}_\perp = \sqrt{\hat{k}_x^2 + \hat{k}_y^2}$ the perpendicular wave number.  \\
Two coefficients are included:
\beq 
\bepes = 8 \pi \frac{n_0 \Tpes }{B_0^2}, \qquad \thea = \frac{\Tpes}{\Tpas},
\eeq
corresponding to the ratio between the perpendicular equilibrium kinetic pressure and the magnetic pressure, and to the temperature anisotropy for each species $s$, respectively.\\
Gyrofluid moments, such as the perturbation of the gyrocenter density,  parallel velocity and of parallel and perpendicular temperature, are computed as moments of the perturbation of the gyrocenter distribution function $\hdfa$. We consider here a model retaining the first four moments, given by
\beq \label{moments}
\hns = \int \dwa \hdfa, \qquad   \frac{\hus}{\vtpa}= \frac{1}{n_0}\int \dwa \frac{\hvpa}{\vtpa}\hdfa,
\eeq
\beq \nonumber
\frac{\hTpas}{\Tpas}=\frac{1}{n_0}\int \dwa \left( \frac{\hvpa^2}{\vtpa^2} -1\right) \hdfa, \qquad 
\frac{\hTpes}{\Tpes}=\frac{1}{n_0}\int \dwa \left( \frac{\hat{\mu}_s B_0}{\Tpes} - 1 \right) \hdfa. %
\eeq
In particular, we will derive evolution equations for the gyrocenter density and parallel velocity fluctuations, but we will need to take into account also gyrocenter temperature fluctuations in order to impose an isothermal closure on the {\it particle} temperature fluctuations. All the fluctuations of the remaining higher order gyrocenter moments are assumed to be zero.

The perturbed gyrocenter distribution function can be developed as a series of its gyrocenter moments using Hermite polynomials $H_n$ and Laguerre polynomials $L_m$, where $H_n$ are polynomials of $\hvpa / \vtpa$ and $L_m$ are polynomials of $ \hat{\mu}_s B_0 / \Tpes$. The expansion reads
\beq
\hdfa(\hbr, \hvpa, \hat{\mu}_s, \hht) = \sum_{m,n=0}^{+ \infty} \frac{1}{\sqrt{m !}}g_{mn_s}(\hbr,\hht) H_m \left( \frac{\hvpa}{\vtpa} \right) L_n\left( \frac{\hat{\mu}_s B_0}{\Tpes} \right) \hcalfa (\hvpar,\hmua).
\eeq
Using the orthogonality relations of  Hermite and Laguerre polynomials, and retaining the first four  moments, the gyrokinetic function $\hdfa$ is written as
\beq  \label{expf}
\hdfa = \hcalfa  \left( \frac{\hns}{n_0} + \frac{\hvpa}{\vtpa} \frac{\hus}{\vtpa} + \frac{1}{2}\left( \frac{\hvpa^2}{\vtpa^2} -1\right) \frac{\hTpas}{\Tpas} + \left( \frac{\hat{\mu}_s B_0}{\Tpes}-1 \right) \frac{\hTpes}{\Tpes} \right).
\eeq
Inserting this expression for $\hdfa$ into Eq. (\ref{1}) and integrating over $\dwa$, gives the evolution equation for $\hns$:
\beq \label{modelgeneral1}
\begin{split} 
\frac{\partial }{\partial \hht} \frac{\hns}{n_0}+& \frac{c}{B_0} \left[ \Guo \hphi + \frac{\Tpes}{q_s} 2 \Gdo \frac{\hbpar}{B_0},  \frac{\hns}{n_0} \right] - \frac{c}{B_0}\left[ \Guu \hphi +  \frac{\Tpes}{q_s}2 \Gdu \frac{\hbpar}{B_0} , \frac{\hTpes}{\Tpes} \right]\ \\
& - \frac{1}{B_0} \left[ \Guo \hapar, \hat{U}_s \right] +  \frac{\partial \hat{U}_s }{\partial \hz} =0.
\end{split}
\eeq
Multiplying (\ref{1})  by $\hvpa /(n_0 \vtpa)$ and integrating over $\dwa$, gives the evolution equation for $\hat{U}_s$:
\beq \label{modelgeneral2}
\begin{split}
\frac{\partial}{\partial \hht}  & \left( \frac{\hat{U}_s}{\vtpa} +  \frac{q_s \vtpa}{\Tpas c} \Guo \hap \right)  + \frac{c}{ B_0} \left[ \Guo \hphi + \frac{\Tpes}{q_s} 2 \Gdo \frac{\hbpar}{B_0}, \frac{\hat{U}_s}{\vtpa} \right] \\ 
& - \frac{\vtpa}{B_0}\left[ \Guo \hap , \frac{\hns}{n_0} + \frac{\hTpas}{\Tpas} \right] + \frac{\vtpa}{B_0}\left[ \Guu \hap ,   \frac{\hTpes}{\Tpes} \right] + \frac{q_s \vtpa}{\Tpas B_0} \sum^{+ \infty}_{n=0}\left[ \mathcal{G}_{1n_s} \hphi, \mathcal{G}_{1n_s}\hap \right]  \\
&+ \frac{\thea\vtpa}{B_0}\sum^{+ \infty}_{n=0} \left[ 2 \mathcal{G}_{2n_s} \frac{\hbpar}{B_0}, \mathcal{G}_{1n_s} \hap \right] +                       \vtpa \frac{\partial}{\partial \hz}\left( \frac{q_s}{\Tpas} \Guo\hphi + 2 \frac{\Tpes}{\Tpas}\Gdo\frac{\hbpar}{B_0} + \frac{\hat{N}_s }{n_0}+ \frac{\hat{T}_{ \parallel s}}{\Tpas}\right) =0,
\end{split}
\eeq
where the operators $\mathcal{G}_{1n_s}$ and $\mathcal{G}_{2n_s}$ are defined in Fourier space so that \cite{Bri92}
\beq\label{g1n}
\mathcal{G}_{1n_s} f (\hx,\hy,\hz )  = \sum_{\hbk}  \frac{B_0}{\Tpes} \int d \hat{\mu}_s \, e^{- \frac{\hat{\mu}_s B_0}{\Tpes}} L_n\left(\frac{\hat{\mu}_s B_0}{\Tpes}\right) J_0(a_s)  f_{\hbk} e^{i \hbk \cdot \hbr} =\sum_{\hbk } \frac{e^{-b_s/2}}{n!}\left( \frac{b_s}{2} \right)^n f_{\hbk} e^{i \hbk \cdot \hbr}  ,
\eeq
\beq
\begin{split}\label{g2n}
\mathcal{G}_{2n_s} f (\hx,\hy,\hz )  & = \sum_{\hbk} \frac{B_0}{\Tpes} \int d \hat{\mu}_s \, e^{- \frac{\hat{\mu}_s B_0}{\Tpes}} L_n\left(\frac{\hat{\mu}_s B_0}{\Tpes}\right) \frac{\hat{\mu}_s B_0}{\Tpes} \frac{J_1(a_s)}{a_s}  f_{\hbk} e^{i \hbk \cdot \hbr}  \\
& = - \sum_{\hbk }  \frac{e^{-b_s/2}}{2}\left( \left( \frac{b_s}{2} \right)^{n-1} \frac{1}{(n-1)!} - \left( \frac{b_s}{2} \right)^n \frac{1}{n!} \right) f_{\hbk} e^{i \hbk \cdot \hbr}   , \quad \text{for} \, n \geq 1,
\end{split}
\eeq
\beq
\mathcal{G}_{20_s} f (\hx,\hy,\hz )   = \sum_{\hbk }  \frac{e ^{-b_s/2}}{2} f_{\hbk} e^{i \hbk \cdot \hbr},
\eeq
with $b_s = \hat{k}_\perp^2 \rho_{th_{\perp_s}}^2$ and $\rho_{th_{\perp_s}}=(1/ \omega_{cs})\sqrt{\Tpes / m_s}$ the perpendicular thermal Larmor radius. 
With regard to the static equations (\ref{2})-(\ref{4}), again, by inserting the expansion (\ref{expf}), we obtain 
\beq \label{quasineutralityappendix}
\sum_s q_s \left( \Guo \frac{\hns}{n_0} - \Guu \frac{\hTpes}{\Tpes} + \frac{q_s}{\Tpes}( \Gamma_{0s} -1) \hphi  + ( \Gamma_{0s} - \Gamma_{1s} ) \frac{\hbpar}{B_0} \right)  =0,
\eeq
\beq \label{ampere1appendix}
- \lapp \hapar = \frac{4 \pi n_0}{c} \sum_s q_s \left( \Guo \hus + \frac{q_s}{m_s}\left( 1 - \frac{1}{\thea}\right) ( \Gamma_{0s} - 1) \frac{\hapar}{c}\right ),
\eeq
\beq \label{ampere2appendix}
\sum_s \bps \left( 2 \Gdo \frac{\hns}{n_0} - 2 \Gdu \frac{\hTpes}{\Tpes} \right) = - \sum_s \bps \frac{q_s}{\Tpes}( \Gamma_{0s} - \Gamma_{1s}) \hphi - 2\frac{\hbpar}{B_0} + \left(\sum_s \bps2(\Gamma_{0s} - \Gamma_{1s} ) \right) \frac{\hbpar}{B_0},
\eeq
where
\beq \label{gammaoperators}
\Gamma_{0s}  f (\hx,\hy,\hz ) = \sum_{\hbk }  I_0(b_s) e^{-b_s}f_{\hbk} e^{i \hbk \cdot \hbr} , \qquad \Gamma_{1s}  f (\hx,\hy,\hz ) = \sum_{\hbk } I_1(b_s) e^{-b_s}f_{\hbk} e^{i \hbk \cdot \hbr} ,
\eeq
and $I_n$ are the modified Bessel functions of order $n$.\\
The system given by Eqs. (\ref{modelgeneral1}), (\ref{modelgeneral2}) and (\ref{quasineutralityappendix}), (\ref{ampere1appendix}), (\ref{ampere2appendix}) requires a closure on the temperature fluctuations. We impose the following standard closure relations consisting in setting the perturbations of the parallel and perpendicular particle temperatures to zero :
\beq \label{tpar}
\frac{\hat{t}_{\parallel_s}}{\Tpas} =  \frac{1}{n_0} \int_0^{2\pi} \frac{d \theta }{2 \pi}\int\dwa \left( \frac{\hvpa^2}{\vtpa^2} -1 \right) \hpdf =0,
\eeq
\beq \label{tperp}
\frac{\hat{t}_{\perp_s}}{\Tpes}=  \frac{1}{n_0} \int_0^{2\pi} \frac{d \theta }{2 \pi}\int  \dwa \left( \frac{\hat{\mu}_s B_0}{\Tpes} -1 \right) \hpdf = 0,
\eeq
where $\hpdf$ is the perturbation of the particle distribution function for a particle of species $s$.
Although the isothermal closure relations (\ref{tpar})-(\ref{tperp}) are easily expressed in terms of the particle temperature fluctuations, in order to include them in our gyrofluid model, we need to express them in terms  of the gyrocenter temperature fluctuations. Therefore we introduce the gyrocenter position $\hat{\textbf{R}} = \hbr + ( \hat{\textbf{v}}/ \omega_{cs}) \times \textbf{z}= \hbr + \hbrhos$, expressed in terms of the particle position $\hbr$ and the particle velocity $\hat{\textbf{v}}=\hvpa \textbf{z} + \hvpe (\cos\theta \bx + \sin\theta \boldy) $, with $\theta = \arctan( v_y / v_x)$ indicating the gyroangle.

To express the closure in terms of the gyrocenter moments, we use the  following relation, which can be obtained from Ref. \cite{kunz2015}:
\beq \label{parttogyro}
\begin{split}
\hat{\mathsf{f}}_{s_{\hbk}} e^{i \hbk. \hbr} = \hat{f}_{s_{\hbk}} e^{i \hbk. \hbr} & + \frac{q_s}{\Tpes} \hcalfa  J_0(a_s)(\hphi_{\hbk}(t) + \frac{\hvpa}{c}(\thea -1) \hat{A}_{\parallel_{\hbk}}(t) )  e^{i \hbk. (\hbr+ \hbrhos)} \\
 & - \frac{q_s}{\Tpes} \hcalfa \left( ( \hphi_{\hbk}(t) + \frac{\hvpa}{c}( \thea -1) \hat{A}_{\parallel_{\hbk}}(t) ) e^{i \hbk.\hbr } \right)\\
 & + \frac{q_s}{\Tpes} \hcalfa \bigg(   2\frac{\hat{\mu}_s B_0}{q_s } \frac{J_1(a_s)}{a_s}\frac{\hat{B}_{\parallel_{\hbk}}(t)}{B_0}e^{i \hbk( \hbr + \hbrhos)} \bigg) \Bigg) =0 .
\end{split}
\eeq
We make use of the relation (\ref{parttogyro}), expressing the perturbation of the particle distribution function $\hpdf$ in terms of that of the gyrocenter $\hdfa$,  into Eqs. (\ref{tpar}) - (\ref{tperp}). By means of the identity
\beq
J_0(a_s) \hat{f}_{s_{\hbk}} = \frac{1}{2 \pi} \int_{0}^{2 \pi } d \theta \hat{f}_{s_{\hbk}} e^{i \hbk \cdot \hbrhos},
\eeq
we obtain that Eqs. (\ref{tpar}) - (\ref{tperp}), lead to the following closure relations in terms of gyrocenter variables:
\beq \label{partemp}
\hat{t}_{\parallel_s} = \Guo \hTpas =0,
\eeq
\beq
\begin{split} \label{perptemp}
\frac{\hat{t}_{\perp_s}}{\Tpes} = - \Guu \frac{\hns}{n_0} + & \left( \Guo -2\Guu + \mathcal{G}_{12_s}  \right)\frac{\hTpes}{\Tpes} -  \frac{q_s}{\Tpes} \mathcal{G}_{T 0_s}  \hphi - \mathcal{G}_{T 1_s}  \frac{\hbpar}{B_0}=0,
\end{split}
\eeq
with the operators
\beq 
\mathcal{G}_{T 0_s} f(\hx,\hy,\hz)= \sum_{\hbk} b_s e^{-b_s}\left( I_0(b_s) - I_1(b_s) \right) f_{\hbk} e^{i \hbk \cdot \hbr},
\eeq
\beq 
\mathcal{G}_{T 1_s} f(\hx,\hy,\hz)= \sum_{\hbk} 2  e^{-b_s}\left( \left( b_s - \frac{1}{2} \right) I_0(b_s) - b_s I_1(b_s) \right) f_{\hbk} e^{i \hbk \cdot \hbr}.
\eeq
We remark that the relations included in Eqs. (\ref{partemp})-(\ref{perptemp}), permitting to express particle temperature fluctuations in terms of gyrocenter temperature fluctuations, agree with those of Ref. \cite{Bri92} (up to a misprint in the sign of $I_1$ in Ref. \cite{Bri92}). In particular, we note that such relations do not depend explicitly on $\Theta_e$.

The resulting gyrofluid system can be conveniently expressed in terms of an appropriate normalization. The adopted dimensionless variables are
\begin{equation} \label{norm2}
\begin{split}
  &  x=\frac{\hx}{\rhos}, \qquad  y=\frac{\hy}{\rhos},  \qquad  z=\sqrt{\frac{\bpe}{2}}\frac{\hz}{\rhos}, \qquad   t=\omega_{ci} \hat{t},  \\
  & N_s=\frac{\hat{N}_s}{n_0},  \qquad    U_s=\sqrt{\frac{\bpe}{2}} \frac{\hat{U}_s}{c_{s \perp}}, \qquad   \tpes = \frac{\hTpes}{\Tpes}, \qquad   \tpas = \frac{\hTpas}{\Tpas},\\
   & \phi=\frac{e \hphi}{T_{0_{\perp e}}}, \qquad   \bpar=\frac{\hbpar}{B_0},  \qquad \apar= \frac{1}{\rhos} \sqrt{\frac{2}{\bpe}} \frac{\hap}{B_0}.\\
   \end{split}  
\end{equation}
A new parameter, naturally emerging from the normalization, is given by
\beq
\taups = \frac{\Tpes}{\Tpee},
\eeq
corresponding to the ratio between the equilibrium perpendicular temperatures. 

The dimensionless evolution equations are given by 
\begin{align} 
&\frac{\partial \ns}{\partial t} +\left[ \Guo \phi,  \ns \right] +   \, \text{sgn}(q_s) \taups  \left[ 2\Gdo \bpara,  \ns \right] - \left[ \Guu \phi, \tpes \right]- \left[ \Guo\apar, \us \right]  \label{nsnorm} \\
&- \text{sgn}(q_s) \taups  \left[ 2 \Gdu \bpara, \tpes \right]   + \frac{\partial U_s}{\partial z} =0,\nonumber\\
&\frac{\partial}{\partial t}  \left( \frac{2}{\bpe}\frac{m_s}{m_i} \us +  \text{sgn}(q_s) \Guo\apar \right)  +  \frac{2}{\bpe} \left[ \Guo \phi , \frac{m_s}{m_i} \us \right] +\text{sgn}(q_s) \taups  \frac{2}{\bpe} \left[ 2\Gdo \bpara, \frac{m_s}{m_i} \us \right]\nonumber  \\
& - \frac{\taups}{\Theta_s}\left[ \Guo\apar , \ns + \tpas \right] + \frac{\taups}{\Theta_s} \left[ \Guu\apar ,   \tpes \right]+ \text{sgn}(q_s)\sum^{+ \infty}_{n=0}\left[ \mathcal{G}_{1n_s} \phi, \mathcal{G}_{1n_s}\apar \right]  \label{ohmslawnorm} \\
&  +\taups\sum^{+ \infty}_{n=0} \left[ 2 \mathcal{G}_{2n_s} \bpara, \mathcal{G}_{1n_s}\apar \right]  + \frac{\partial}{\partial z}\left( \text{sgn}(q_s) \Guo\phi + 2 \taups\Gdo\bpar + \frac{\taups}{\Theta_s}\left( {N}_s  + \tpas \right) \right) =0, \nonumber 
& 
\end{align}
and the static equations correspond to
\begin{align} 
    & \sum_s \text{sgn}(q_s) \left( \Guo \ns - \Guu \tpes +  \frac{\text{sgn} (q_s)}{\taups} ( \Gamma_{0s} -1) \phi + ( \Gamma_{0s} - \Gamma_{1s}) \bpara \right) =0,  \label{contnorm}\\
    & \sum_s \left( \frac{\bps}{2} \frac{2}{\bpe}\text{sgn}(q_s) \Guo \us + \left(1 - \frac{1}{\Theta_s} \right) (\Gamma_{0s} -1)\frac{m_i}{m_s} \frac{\bps}{2 }\apar \right)= - \lapp\apar, \label{ampere1norm}\\
    &  \sum_s \bps \frac{\text{sgn}(q_s)}{\taups}(\Gamma_{0s} - \Gamma_{1s}) \phi + 2\bpara + 2 \sum_s \bps(\Gamma_{0s} - \Gamma_{1s}) \bpara = - 2  \sum_s\bps \left(  \Gdo \ns - \Gdu \tpes \right), \label{ampere2norm}\\
    & \left( \Guo -2\Guu + \mathcal{G}_{12_s}  \right)\tpes - \Guu \ns - \mathcal{G}_{T 0_s}   \frac{\text{sgn} (q_s)}{\taups} \phi - \mathcal{G}_{T 1_s}  \bpara = 0 \label{tperpnorm},\\
    & \Guo \tpas = 0. \label{tparnorm}
\end{align}
The system (\ref{nsnorm})-(\ref{tparnorm}) is a new gyrofluid model accounting for parallel magnetic perturbations, FLR effects, equilibrium temperature anisotropies, and closed by imposing that the particle temperature fluctuations be zero. 

Because of the complexity of the model, for the purpose of an analytical investigation of the tearing instability, we are required to simplify it by applying a number of assumptions.

In particular, we aim at reducing the model (\ref{nsnorm})-(\ref{tparnorm}) to a two-field model consisting of the evolution equations for the fluctuations of the electron  density and electron parallel velocity. 

First, we assume all the involved fluctuations of the ion gyrocenter moments, i.e. $N_i$, $U_i$, $T_{\perp_i}$ and $T_{\parallel_i}$, to be negligiblein the static relations, which effectively decouples the electron dynamics from the ion gyrocenter dynamics. This is of course a very strong assumption, which ignores the evolution of ion gyrocenter density and parallel velocity based on Eqs. (\ref{nsnorm}) and (\ref{ohmslawnorm}) for $s=i$. On the other hand, at least in the case of initial conditions $N_i=U_i=0$, ion gyrocenter density and parallel velocity fluctuations appear not to substantially modify the reconnection process \cite{Com12}. We also assume an isotropic ion temperature, i.e. $\Theta_i=1$. 

We further reduce the system by considering the following ordering,  where $\bpe$, $\de$ and $\tpi$ are used as expansion parameters,
\begin{align}\label{ord2}
  &  \partial_x \sim \partial_y \sim \Theta_e  = O(1), \\ 
 &  \partial_t \sim \partial_z \sim N_e \sim \phi \sim \apar \sim U_e = O( \varepsilon ) \ll 1,  \label{ord2b}\\
&\bpar \sim \tpee  = O(\varepsilon \bpe), \label{ord2c}\\
& \delta^2 \ll \bpe \ll 1,  \label{ord1} \\
&  \tpi \ll 1. \label{ord3}
\end{align}
The ordering (\ref{ord2}) fixes equal to $\rhos$ the characteristic scale length for the variations of the fluctuations in the perpendicular plane, and assumes the electron temperature anisotropy to remain finite as the expansion parameters tend to zero.  The ordering (\ref{ord2b}), on the other hand, refers to small amplitude, low-frequency and strongly anisotropic  fluctuations, which are typical assumptions of the $\delta f$ gyrokinetic approach (note, on the other hand, that due to our subsidiary ordering in the parameter $\bpe$, the ordering (\ref{ord2b}) is not equivalent to the one assumed in Ref. \cite{kunz2015} for deriving the parent gyorkinetic model (\ref{1})-(\ref{4})). All the lowest order terms in the gyrofluid equations (\ref{nsnorm})-(\ref{ohmslawnorm}) are of order $\varepsilon^2$ and no further expansion will be performed in the parameter $\varepsilon$ (not to be confused with the parameter $\epsilon$ introduced in Sec. \ref{ssec:inner} to rescale the inner variable). According to Eq. (\ref{ord1}), we consider a small value of the $\bpe$ parameter, although much larger than the mass ratio $\de^2$. The reason for the latter ordering is that it will allow to neglect electron FLR effects, while retaining electron inertia term in Ohm's law, which is required for reconnection. As a consequence, the fluctuations $\bpar$ and $\tpee$ will turn out to be subdominant with respect to the other fluctuations, as indicated by Eq. (\ref{ord2c}).  
We consider in (\ref{ord3}) a cold ion regime, assuming that the value of the ion to electron perpendicular equilibrium temperature ratio be small. In particular, assuming $\tpi \ll \de^2$ as $\de \rightarrow 0$, effectively removes all finite ion temperature effects, except for the one associated with the ion polarization in the quasi-neutrality relation (\ref{contnorm}), where $\tpi$ appears at the denominator. \\
We focus, in the coming steps, on the reduction of the gyroaverage operators using the assumptions (\ref{ord1}) and (\ref{ord3}). The orderings (\ref{ord1})-(\ref{ord3}) indicate that the normalized parameter  $b_s$ for each species, corresponding to $b_e =  \delta^2 \kp^2$ and $b_i= \tpi \kp^2$, can be expanded, allowing us to simplify the gyroaverage operators $\mathcal{G}_{1n_s}$ and $\mathcal{G}_{2n_s}$, whose general form is given by Eqs. (\ref{g1n}) and (\ref{g2n}). We show below the expression of the operators acting on the electron moments that will be considered for the reduction (analogous expressions apply in the ion case):
\beq\label{ope}
\begin{split}
& \Guoe f(x,y,t) = \sum_{\bk } e^{-b_e/2} f_{\bk} e^{i \bk \cdot \br}, \qquad \Guue f(x,y,t) = \sum_{\bk } \frac{b_e}{2} e^{-b_e/2} f_{\bk} e^{i \bk \cdot \br}, \\
& \Gdoe f(x,y,t)= \sum_{\bk } \frac{ e^{-b_e/2}}{2} f_{\bk} e^{i \bk \cdot \br}, \qquad \Gdue f(x,y,t) = - \sum_{\bk } \frac{ e^{-b_e/2}}{2}\left( 1 - \frac{b_e}{2} \right) f_{\bk} e^{i \bk \cdot \br}. 
\end{split}
\eeq
Upon the ordering (\ref{ord1}), these operators are written using their Taylor expansion as
\beq
\begin{split} \label{approxG}
& \Guoe f(x,y) = \left(1 + \frac{1}{2}\delta^2 \lapp\right) f(x,y) + O(\delta^3), \qquad \quad \Guue f(x,y) =  - \frac{\delta^2 \lapp}{2} f(x,y) + O(\delta^3),\\
&\Gdoe f(x,y) = \frac{1}{2}\left( 1 + \frac{1}{2}\delta^2 \lapp \right) f(x,y) + O(\delta^3), \qquad \Gdue f(x,y) = - \frac{1}{2}\left( 1 + \frac{1}{2}\delta^2 \lapp\right) f(x,y) + O(\delta^3),
\end{split}
\eeq 
while the operators $\mathcal{G}_{1n_e}$ and $\mathcal{G}_{2n_e}$ with $n \geq2$  are of order $O(\de^2)$ and thus turn out to be negligible. Regarding the ion and electron $\Gamma_{ns}$ operators, present in Eqs. (\ref{contnorm}) - (\ref{tparnorm}) and whose general expression is given in Eq. (\ref{gammaoperators}), they can  be simplified as well under Eq. (\ref{ord1}) and written as
\beq \label{approxgamma}
\begin{split} 
& \Gamma_{0e} f(x,y) = ( 1 + \delta^2 \lapp) f(x,y) + O(\delta^3), \qquad  \Gamma_{1e} f(x,y)= O(\de^2), \\
& \Gamma_{0i} f(x,y) = ( 1 + \tpi \lapp) f(x,y) + O(\tpi^2), \qquad \Gamma_{1i} f(x,y)=O(\de^2).
\end{split}
\eeq
Using the ordering (\ref{ord2}) in the closure equations (\ref{tparnorm}) and (\ref{tperpnorm}) and neglecting terms proportional to $\de^2$ compared to terms of order one, we obtain the reduced closure equations, 
\beq\label{closure}
\tpae =0, \qquad \tpee = - \bpar.
\eeq
When applying the ordering (\ref{ord2}) to the evolution equations (\ref{nsnorm}) and (\ref{ohmslawnorm}), the assumption $\de^2 \ll \bpe \ll 1$ allows us to neglect terms proportional to $\delta^2$, arising from the operators $\mathcal{G}_{1n_e}$ and $\mathcal{G}_{2n_e}$, when compared to terms proportional to $\delta^2/\bpe$ and, therefore, to neglect electron FLR effects. However, retaining first order corrections in $\bpe$ allows us to keep some terms involving the perturbation $\bpar$. \\
Therefore, we can write the evolution equations, retaining the above mentioned corrections as well as the subdominant term $(2 / \bpe)[ \bpar, \delta^2 U_e ]$ of order $\delta^2$, allowing the system to keep a Hamiltonian formulation (a similar inconsistency in the ordering, possessing, on the other hand, the merit of preserving the Hamiltonian character of the parent gyrokinetic model, was discussed in Ref. \cite{Pas18}):  
\beq \label{systham}
\frac{\partial N_e}{\partial t} + [\phi - \bpar, N_e] - [\apar, U_e] + \frac{\partial U_e}{\partial z}=0,
\eeq
\beq \label{systham2}
\frac{\partial }{\partial t}\left( \apar -  \frac{2\de^2}{\bpe}U_e\right) + \left[\phi - \bpar, \apar  -  \frac{2\de^2}{\bpe}U_e \right] + \frac{1}{\Theta_e} [ \apar , N_e] + \frac{\partial}{\partial z}\left( \phi - \bpar - \frac{N_e}{\Theta_e}\right)=0. 
\eeq
Applying now the ordering to the static equations (\ref{contnorm}), (\ref{ampere1norm}), (\ref{ampere2norm}) and neglecting terms proportional to $\delta^2$,  while retaining first order corrections in $\bpe$, yields the relations 
\beq \label{staticequationsappendix}
N_e = \lapp \phi, \qquad U_e = \left( 1 + \frac{\bpe}{2}\left( 1 - \frac{1}{\Theta_e} \right) \right) \lapp\apar, \qquad \bpar = - \frac{\bpe}{2+\bpe} \lapp \phi. 
\eeq
By means of the first relation in  Eq. (\ref{staticequationsappendix}), indicating that the electron gyrocenter density equals the $\mathbf{E}\times \mathbf{B}$ vorticity $\lapp \phi$, Eq. (\ref{contnorm}) becomes an evolution equation for the vorticity. \\
The system (\ref{systham})-(\ref{systham2})  was shown to be Hamiltonian in Ref. \cite{Tas19}. Its Hamiltonian structure consists of the Hamiltonian functional
\begin{equation}
    H(N_e, A_e)= \frac{1}{2} \int d^3 x \, \left( \frac{N_e^2}{\Theta_e} - A_e \lapp \lba A_e  -N_e \lbphi N_e +N_e \lbb N_e  \right),
    \label{eq:ham}
\end{equation}
and of the Poisson bracket 
\begin{equation}
\begin{split}
     & \{ F , G \}= \int  d^3 x \, \Bigg( N_e \left(  [F_{N_e}, G_{N_e}]  + \frac{\delta^2}{\Theta_e}[F_{A_e} , G_{A_e}]\right)  \\
     & + A_e([F_{A_e} , G_{N_e}] + [F_{N_e} , G_{A_e}]) + F_{N_e} \frac{\partial G_{A_e}}{\partial z} + F_{A_e} \frac{\partial G_{N_e}}{\partial z} \Bigg).
    \label{eq:pb}
\end{split}
\end{equation}
In Eqs. (\ref{eq:ham}) and (\ref{eq:pb}) $A_e=\apar - 2\delta^2  U_e /\bpe$, whereas $\lba$, $\lbphi$ and $\lbb$ are linear operators that permit to express $\apar$, $\phi$ and $\bpar$ in terms of $N_e$ and $A_e$ by means of Eqs. (\ref{staticequationsappendix}). 
Using the relations (\ref{closure}) and (\ref{staticequationsappendix}), the evolution equations (\ref{systham}) and (\ref{systham2}) become
\beq \label{continuitywithbpe}
\begin{split}
\frac{\partial \lapp \phi}{\partial t} + [\phi, \lapp \phi] - & \left( 1 + \frac{\bpe}{2}\left( 1 - \frac{1}{\Theta_e} \right) \right)[\apar, \lapp \apar] \\
& + \frac{\partial}{\partial z}\left( 1 + \frac{\bpe}{2}\left( 1 - \frac{1}{\Theta_e} \right) \right) \lapp\apar=0,
\end{split}
\eeq
\begin{equation}
\begin{split} \label{ohmslawwithbpe}
\frac{\partial}{\partial t}\Bigg( \apar -  \frac{2\de^2}{\bpe} & \left( 1 +  \frac{\bpe}{2}\left( 1 - \frac{1}{\Theta_e} \right) \right) \lapp\apar\Bigg) \\ 
& + \left[\phi, \apar  -  \frac{2\de^2}{\bpe} \left( 1 + \frac{\bpe}{2}\left( 1 - \frac{1}{\Theta_e} \right) \right) \lapp \apar \right] \\
& - \frac{\bpe}{\bpe +2 }\left[\lapp \phi, \apar  -  \frac{2\de^2}{\bpe} \left( 1 + \frac{\bpe}{2}\left( 1 - \frac{1}{\Theta_e} \right) \right) \lapp \apar \right] \\
& + \frac{1}{\Theta_e} [\lapp \phi, \apar] + \frac{\partial}{\partial z}\left( \phi  + \frac{\bpe}{2+\bpe} \lapp \phi - \frac{\lapp \phi}{\Theta_e}\right)  =0.
\end{split}
\end{equation}
As a further simplification, by virtue of Eq. (\ref{ord2}), one can neglect terms of order $\bpe$ when compared to terms of order unity, which leads to  the system
\beq  \label{continuityappendix}
\frac{\partial \lapp \phi}{\partial t} + [\phi, \lapp \phi] - [\apar, \lapp \apar] + \frac{\partial \lapp \apar}{\partial z}=0,
\eeq

\begin{equation}
\begin{split}  \label{ohmslawappendix}
\frac{\partial}{\partial t} \left( \apar -  \frac{2\de^2}{\bpe} \lapp \apar \right) + \left[\phi, \apar  -  \frac{2\de^2}{\bpe} \lapp \apar \right]  - \frac{1}{\Theta_e} [\lapp \phi, \apar] + \frac{\partial }{\partial z}\left( \phi - \frac{\lapp \phi}{\Theta_e} \right)=0.
\end{split}
\end{equation}
Eqs. (\ref{continuityappendix}) and (\ref{ohmslawappendix}) are indeed those considered for our tearing stability analysis and correspond to Eqs. (\ref{continuity})-(\ref{ohmslaw}). The analysis can easily be extended to account also for the small modifications of the coefficients due to the presence of a finite $\bpe$. This leads to the generalized dispersion relation (\ref{grlargebeta}).

\section{Convergence of a limit relevant for the outer solution $\wphi_{out}$ }  \label{app:lim}

 In this Appendix we show that
\beq 
\begin{split} \label{intergal}
\lim_{x \rightarrow + \infty} e^{- \alpha x} \int_{a}^{x} \left( \frac{1}{k_y} + \frac{\lambda}{\tanh (t/\lambda)} \right) e^{( \alpha - k_y) t }dt =0,
\end{split}
\eeq
which is necessary in order to verify that the solution (\ref{outersolutionphi}), satisfies the boundary condition $ \lim_{x \rightarrow + \infty} \wphi_{out} = 0$. 

First, we recall  that the coefficient $\alpha - k_y = \sqrt{k_y^2 + \Theta_e} - k_y$ is positive. \\
Then we use the fact that $| 1/ \tanh t| < |1/t + 1|$ on the domain $t \in ]0, + \infty) $. This yields 
\beq
\begin{split}
& 0 \leq e^{- \alpha x}  \int_{a}^{x}  \left( \frac{1}{k_y} + \frac{1}{\tanh (t/\lambda)} \right) e^{ (\alpha -k_y )t}dt <     e^{- \alpha x}    \int_a^{x} \left( \frac{1}{k_y} + \frac{\lambda}{t} + 1 \right)e^{ (\alpha - k_y)t}  dt  \\
&  = e^{- \alpha x} \Bigg(  \lambda\int_{(\alpha - k_y)a}^{(\alpha-k_y)x} \frac{e^u}{u}du  + \frac{1/k_y +1 }{k_y -\alpha} \Big[ e^{ (\alpha-k_y) a } - e^{(\alpha-k_y)x} \Big]\Bigg) \\
& = e^{- \alpha x} \Bigg(  \lambda  E_i\big( (\alpha- k_y)x \big) -  \lambda E_i\big( (\alpha- k_y)a\big)  +\frac{1/k_y +1 }{k_y -\alpha} \Big[ e^{  (\alpha-k_y)a } - e^{(\alpha-k_y)x} \Big]\Bigg)\\
&  \underset{x \rightarrow +\infty}{\sim} e^{- k_y x} \left( \frac{\lambda}{(\alpha - k_y) x} + O\left(\frac{1}{x^2}\right) \right) - \lambda e^{- \alpha x} Ei((\alpha-k_y)a)  + \frac{1/k_y +1 }{k_y -\alpha} \left( e^{- \alpha x + ( \alpha - ky)a} - e^{-k_y x} \right)\\
& \underset{x \rightarrow +\infty}{\rightarrow} 0,
\end{split}
\eeq
where $Ei$ is the exponential integral function and where in the last step we made use of the asymptotic expansion $Ei(x) \sim e^{-x}\left( \frac{1}{x} + \frac{1}{x^2} + O (\frac{1}{x^3}) \right) $.
This shows the convergence of the integral (\ref{intergal}).

\end{appendix}

\bibliographystyle{plain} 
\bibliography{biblio.bib}

\end{document}